\documentclass[reprint,amsmath,amssymb,aps,prl]{revtex4-2}

\usepackage{graphicx}
\usepackage{dcolumn}
\usepackage{bm}
\usepackage{physics}
\usepackage{mathtools}
\usepackage{relsize}
\usepackage{bigints}
\usepackage{helvet}
\usepackage[colorlinks=true, allcolors=blue]{hyperref}
\usepackage{academicons}
\usepackage{amsmath}
\usepackage{amsbsy}
\usepackage{lipsum}
\usepackage{color}
\usepackage{tikz,xcolor}
\usepackage{placeins}
\usepackage[caption=false]{subfig}
\usepackage{array} 
\usepackage{booktabs} 
\usepackage{adjustbox} 
\usepackage{relsize}
\usepackage{multirow}
\usepackage{qtree}
\usepackage{forest}
\usepackage{tikz}
\usetikzlibrary{trees,arrows}
\begin{document}

\title{An Imprecise Maxwell's Demon with Feedback Delay: An Exactly Solvable Information Engine Model}

\author{Kiran V}
\email{kiran.vktm@gmail.com, p20180029@goa.bits-pilani.ac.in}
\affiliation{Department of Physics, BITS Pilani K K Birla Goa Campus, 
Zuarinagar 403726, Goa, India}
\author{Toby Joseph}
\email{toby@goa.bits-pilani.ac.in}
\affiliation{Department of Physics, BITS Pilani K K Birla Goa Campus, 
Zuarinagar 403726, Goa, India}

\begin{abstract}
A finite cycle time information engine based on a two-level system in contact with a thermal reservoir is studied analytically. The model for the engine incorporates an error in measuring the system's state and time delay between the measurement and the feedback process. The efficiency and power of the engine in steady state are derived as a function of level spacing, feedback delay time, engine cycle time, and measurement error. For a fixed value of level spacing and feedback delay, there is an upper bound on measurement error such that the engine can extract positive work. This threshold value of error is found to be independent of the cycle time. For a range of values of level spacing and feedback delay time, efficiency has a non-monotonic dependence on the measurement error, implying that there is an optimal measurement error for the information engine to operate efficiently. At high temperatures and with precise measurement, the engine's ability to extract positive work is extended over a larger range of feedback delay time.
\end{abstract}

\maketitle
Maxwell's demon-like setups facilitate heat extraction from a thermal bath and its conversion into useful work through feedback control \cite{Maruyama2009}. In the Szilard engine implementation of the Maxwell demon concept, the feedback control operates as follows: the demon measures whether a single molecule within a vessel, in contact with the thermal bath, is located on the left or right half of the vessel. This information is then utilized to extract work by inserting a partition into the box and a subsequent isothermal expansion of the volume containing the particle \cite{szilard1964decrease}. 
In the first analysis, this engine seems to violate the second law of thermodynamics. However, after nearly half a century of debates and discussions, it is now agreed that work can be extracted from such a system without contravening the second law of thermodynamics as long as the energy cost of the information processing performed by the demon is duly considered \cite{Landauer1961,bennett1982thermodynamics,Maruyama2009,sagawa2009minimal,mmaxwell} (For alternative perspectives on this debate, refer to the following sources \cite{earman1998exorcist,earman1999exorcist,hemmo2010maxwell,norton2011waiting,kish2012energy}). Commonly referred to as information engines, they have been experimentally realized in various classical \cite{toyabe2010experimental,berut2012experimental,saha2021maximizing,paneru2018lossless} and quantum \cite{PhysRevLett.113.030601,
PhysRevResearch.2.032025,averin2011maxwell, camati2016experimental,
naghiloo2018information,chida2017power} systems. 

Two modifications can be considered for a practical engine assessment:
(i) introducing a delay between measurement and feedback, and (ii) addressing errors in the measurement process. Regarding the Szilard engine version of Maxwell's demon, a feedback delay time implies a time lag between measuring the particle's position in the box and inserting the partition. In any experimental implementation of Maxwell's demon, it is a challenging task to achieve instantaneous feedback following the measurement \cite{saha2021maximizing}. Therefore, accounting for the delay between measurement and feedback in a practical information engine context is essential. Given this experimental constraint, it becomes imperative to investigate how the engine's capacity for work extraction and efficiency depends on the feedback delay. Such an analysis helps determine the permissible range of feedback delay for the 
system to effectively utilize the information and operate as an engine 
\cite{toyabe2010experimental,PhysRevLett.113.030601}. As with feedback delay, measurement errors are unavoidable in practice \cite{dinis2020extracting, Um_2015,Hoppenau_2014}. 
Moreover, there is a cost associated with running more precise engines, which has to be kept in mind while optimizing the efficiency of the engine as a whole. Thus, it is important to see how the accuracy of the measurement influences the information engine's 
performance parameters.

Many recent works, both in experiments \cite{saha2021maximizing,paneru2018optimal,rico2021dissipation} and in theory \cite{lucero2021maximal,dinis2020extracting,pal2014extracting} have looked at ways to improve the efficiency and power of information engines. Information engines based on colloidal particles moving through a harmonic potential \cite{abreu2011extracting, bauer2012efficiency, paneru2018lossless} and periodic potentials \cite{toyabe2010experimental, PhysRevE.106.054146} have been studied. 
These studies look into the possibility of extracting work or converting the information about the particle's position into work with the help of a feedback scheme. The simplest model that can be studied to investigate the non-equilibrium dynamics of an information engine is based on a two-level system. The efficiency and power of such a two-level information engine without feedback delay have been studied analytically in the limit of infinite cycles and relaxation time \cite{Um_2015}. An analytical study of a similar engine based on a two-level system incorporating feedback delay but without error has recently been done \cite{kiran3}. The present work is an analytic study of the most general information engine of this type, which incorporates both feedback delay and error in measurement. Further, the results are derived for the engine working with a finite cycle time rather than assuming equilibrium at the beginning of each cycle. Under the assumption of a non-equilibrium steady state, the power and efficiency of the engine are derived, and their dependence on cycle time, feedback delay time, measurement error, and the energy difference between the levels is studied.

{\it The model:}
The information engine consists of a system with two states, with energies $+U_0$ 
(the up-state) and $-U_0$ (the down-state). The system is in contact with a thermal reservoir at temperature $T$. In addition, a demon measures the system's state at 
regular intervals of time, $t=n\alpha$ ($n$ an integer), where $\alpha$ is the cycle time. The demon initiates feedback depending on the outcome of the measurement done. The feedback protocol follows: If the system is measured in the up-state, the demon switches the system's state to the down-state at a time $t = n\alpha + \epsilon$. $\epsilon$ is the feedback delay time and is less than the cycle time, $\alpha$. If the system is measured to be in the down-state during the measurement, the demon carries out no feedback process.

The measurement carried out by the demon is prone to error. That is, there is a 
chance that during the measurement process, the demon measures the state to be 
down when the actual state of the system is up and vice versa. We assume the error 
made by the demon to be such that the conditional probabilities 
$P(M = d| X = u) = \delta_a$ and $P(M = u | X = d) = \delta_b$, where variable $X$ 
represents the actual state of the system, and variable $M$ represents the measurement 
outcome. $u$ and $d$ denote the up- and down-states, respectively. It is assumed that 
$\delta_a = \delta_b \equiv \delta$ for simplicity. It follows that 
$P(M = d | X = d) = P(M = u | X = u) = 1 - \delta$. It is to be noted that 
the feedback will be initiated by the demon depending on the measurement outcome 
and not on the actual state of the system.

The master equation for the process is given by
\begin{equation}
\frac{d p_u}{dt} = -(k_1 + k_2) p_u + k_2\;,
\end{equation}
where $p_u$ is the probability of finding the system in the upstate. $k_1$ is the transition rate from up-state to down-state, and $k_2$ is the transition rate from down-state to up-state. The probability of being in the down-state is given by $p_d = 1 - p_u$. 
Detailed balance in equilibrium stipulates $\frac{k_1}{k_2} = e^{2\beta U_0}$. The relaxation time for the system to reach equilibrium starting from an arbitrary initial state is given by $\tau = \frac{1}{k_1 + k_2}$. When the measurement outcome is up-state, the master equation has to be integrated in two time segments: from $t = n \alpha$ to $t = n \alpha + \epsilon$ and then from $t = n \alpha + \epsilon$ to $t = (n + 1) \alpha$. This is because the state will be switched after a delay time of $\epsilon$ for this measurement outcome. The information engine will reach a steady state after many cycles of operation. The steady state probability is derived below. This allows one to obtain the average work extracted and the average information processed during the measurement, using which efficiency and power of the information engine are found. 

{\it The steady state solution:} Let $p_{u}^{ss}(t) $ be the steady state probability of the system to be in the up state at time $t$. Then, for $\epsilon < t \le \alpha$,
\begin{equation}
\begin{aligned}
p_{u}^{ss}(t) &= \sum_{X'} \biggl[P_2(u; t | X'; 0) P_0(X',d) \biggr] \\
&+ \sum_{X', X''} \biggl[ P_2(u; t | X'';\epsilon^{+}) P_1(X''|X') P_0(X',u) \biggr] \; ,
\end{aligned}
\label{probT}
\end{equation}
where $P_0(X, M)$ is the joint probability for $X$ (state) and $M$ (measurement outcome) 
at the time of measurement ($t = 0$). $P_1(X'' | X')$ is the probability for the state 
to be $X''$ at $t = \epsilon^+$ (that is, right after feedback), given that the state was $X'$ at $t = 0$ and the measurement outcome was $M = u$. 
$P_2(X\; ; t_2\;| X'\; ; t_1)$ is the probability to find the system is in state 
$X$ at $t = t_2$ given that it was in $X'$ at $t = t_1$ with no interference 
from the demon in the interval between $t_1$ and $t_2$ and is given by,
\begin{equation}
 \text{\smaller$ P_2(X;t_2|X';t_1) = p^{eq}(X')(1 -  e^{-(t_2 - t_1)/\tau}) 
+ \tilde{\delta}_{X X'} \;e^{-(t_2 - t_1)/\tau}  \;$}.
\label{P2}
\end{equation}
\noindent
Here $\tilde{\delta}_{X X'}$ is the Kronecker delta function and 
$p^{eq}(X')$ is the equilibrium distribution without feedback. For $X = u$,
\begin{equation}
p^{eq}(X = u) = \frac{e^{-\beta U_0}}{2\cosh{(\beta U_0} )} \equiv p^{eq}_u \;,
\label{eq_prob}
\end{equation}
and $ p^{eq}(X = d) = 1 - p^{eq}_u \equiv  p^{eq}_d$. 
Table I in the supplement gives the set of various joint and conditional probabilities appearing 
in Eq. (\ref{probT}), where $p_{u0}^{ss} \equiv p_{u}^{ss}(t = 0)$ and $p_{d0}^{ss} \equiv 1 - p_{u0}^{ss}$. 

\begin{figure}
    \subfloat[]{\includegraphics[width=0.50\columnwidth]{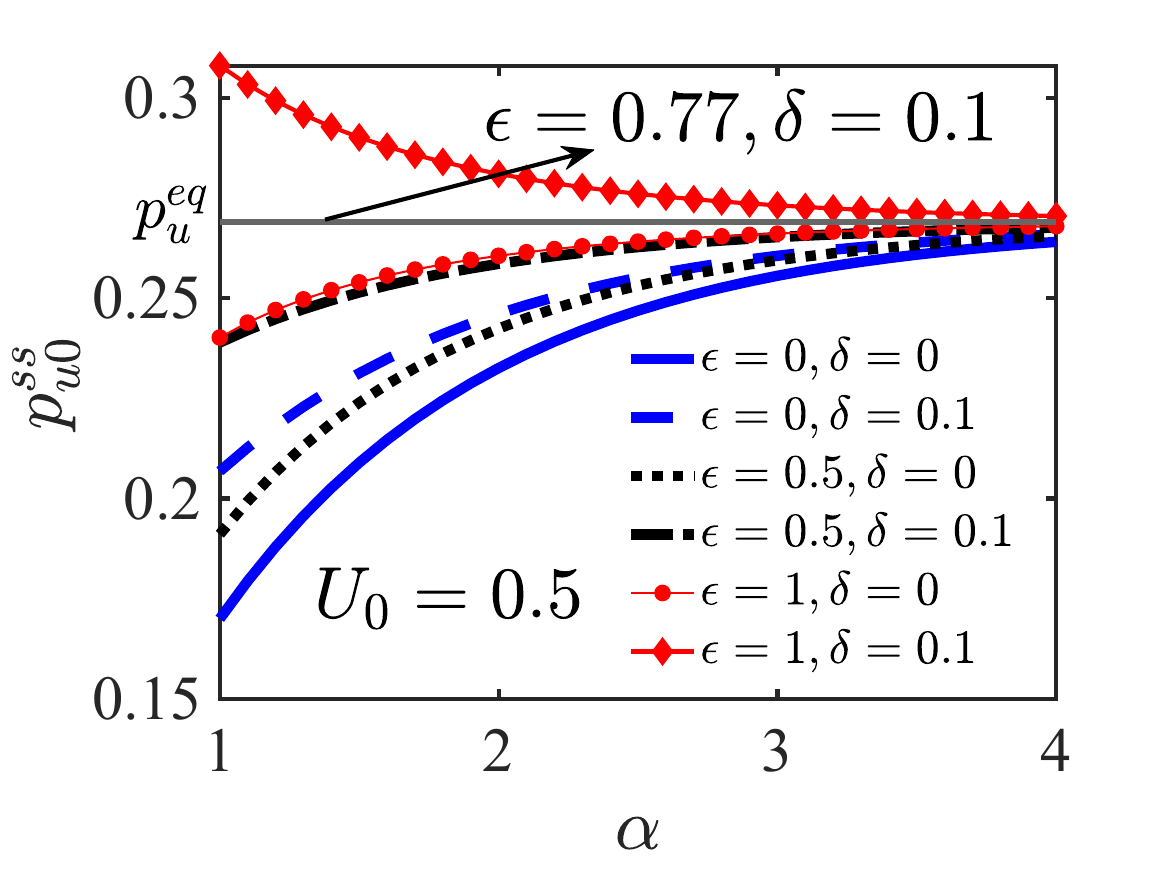}\label{fig:ss}}%
    \subfloat[]{\includegraphics[width=0.50\columnwidth]{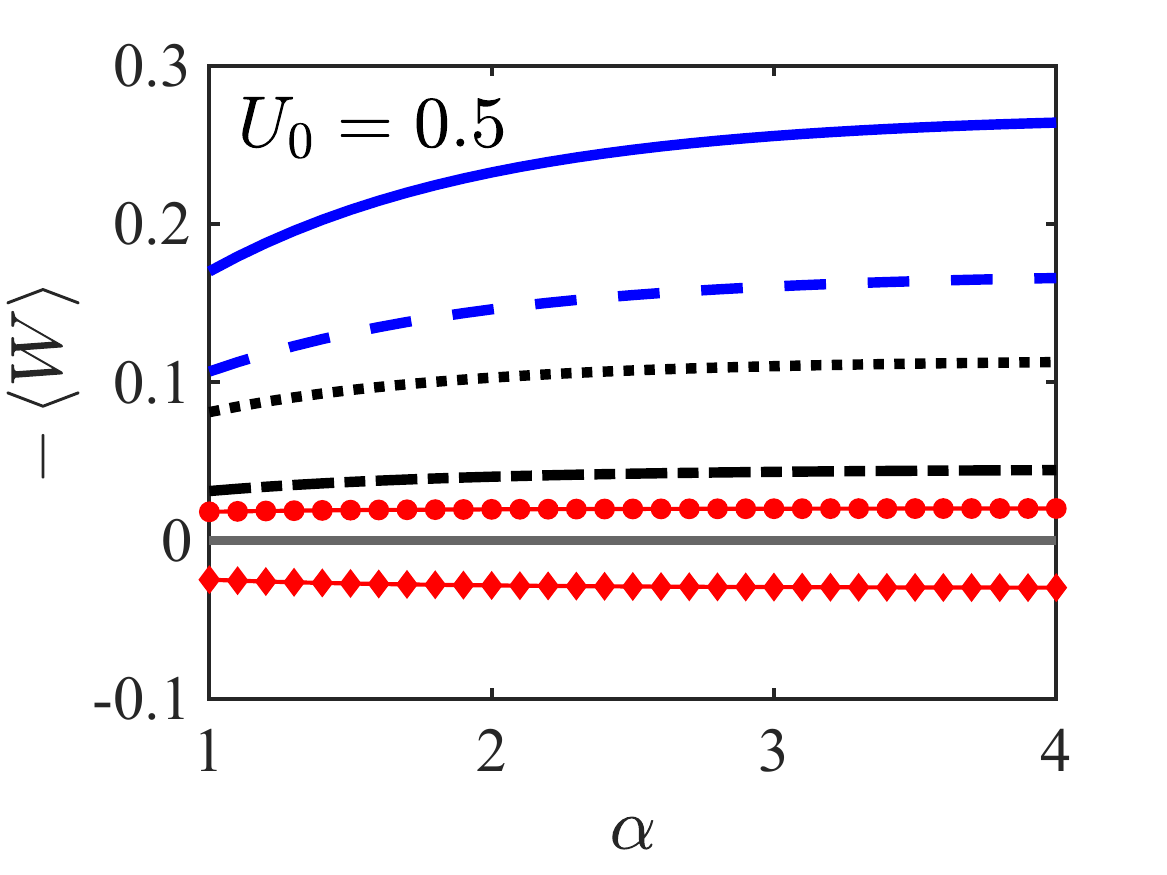}\label{fig:work}}\\
    \subfloat[]{\includegraphics[width=0.50\columnwidth]{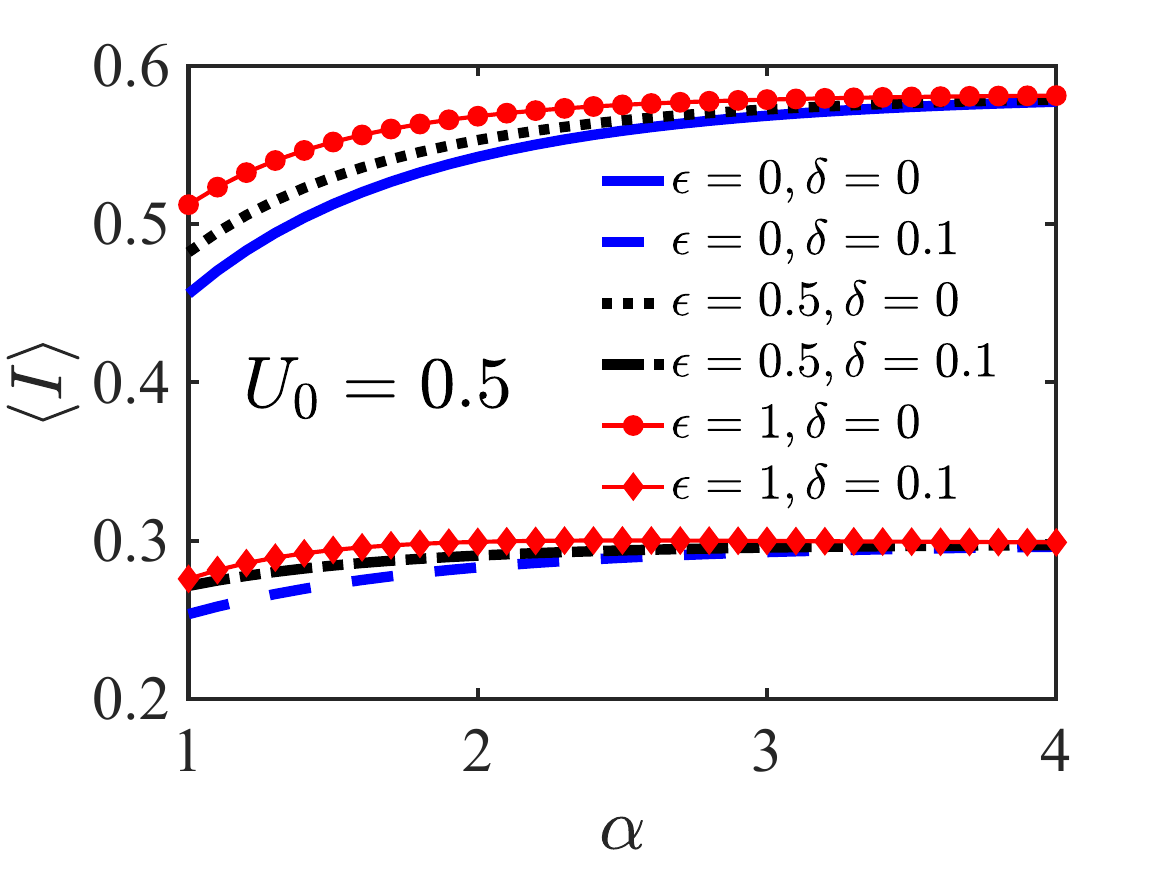}\label{fig:info}}%
    \subfloat[]{\includegraphics[width=0.50\columnwidth]{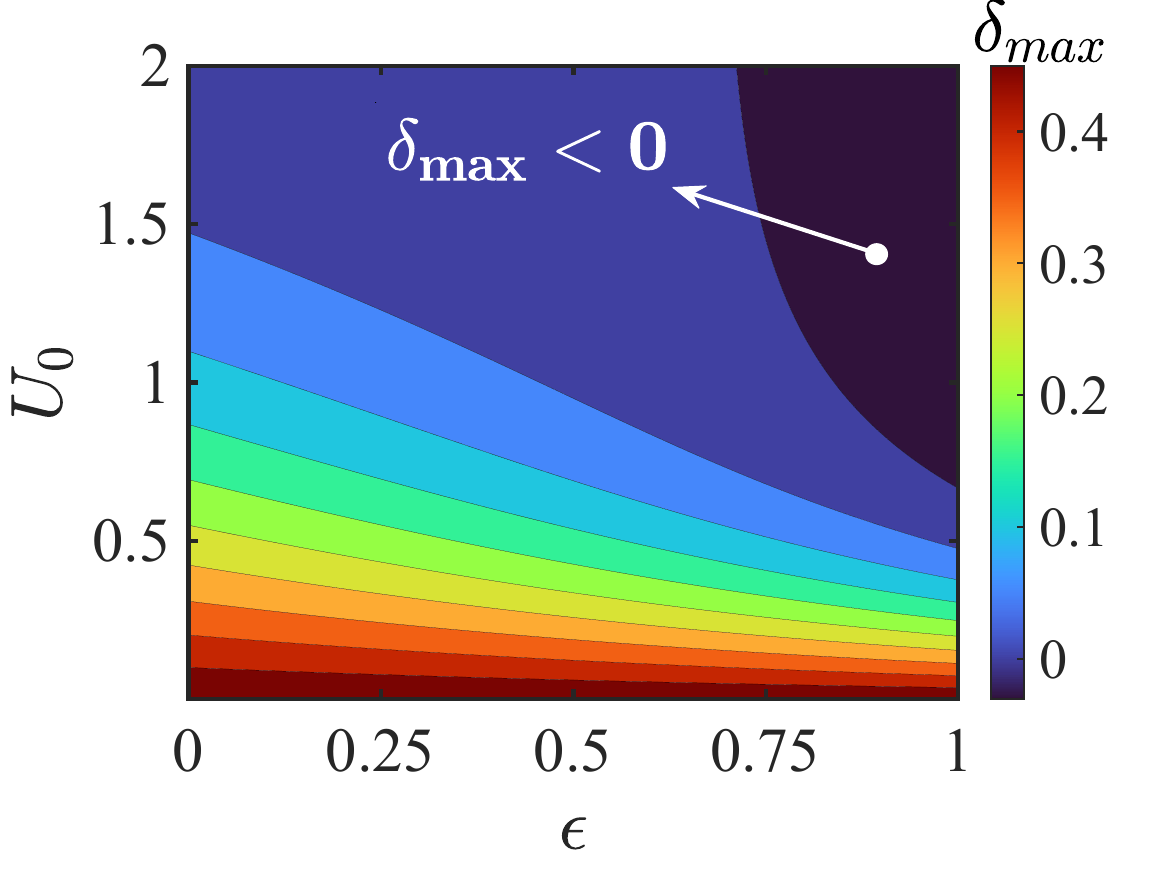}\label{fig:delta_max}}
    \caption{(a) The probability in steady state for the system to be in the up-state 
    at the time of measurement as a function of the cycle time ($\alpha$). For the parameters given, 
    $p_u^{eq} = 0.26$. Immediate feedback and a small measurement error lead to values of 
    $p_{u0}^{ss} < p_u^{eq}$. For a given $\delta$, one can tune $\epsilon$ such that 
    $p_{u0}^{ss} = p_u^{eq}$ and independent of $\alpha$ (black solid horizontal line). 
    (b) Variation of work extracted per cycle as a function of cycle time. Work extraction increases with cycle time and saturates for values larger than the relaxation time. The legends are the same as in (a). (c) Average information as a function of cycle time. There is a significant drop in information gathered with the increase in error. (d) The contour plot gives the maximum value of $\delta$ in the $U_0 - \epsilon$ plane such that the engine can extract positive work. The dark region in the top right corner has negative values, indicating that the engine cannot extract positive work with those parameters even with perfect measurement.}
\label{fig:main1}
\end{figure}

Since $p_{u}^{ss}(t)$ is a periodic function 
with period $\alpha$,  $p_{u0}^{ss}$ must be the same as $p_{u}^{ss}(\alpha)$.
Using this fact in Eq. (\ref{probT}), one gets a self-consistent relation for 
$p^{ss}_{u0}$ (see the Supplement). Solving for $p^{ss}_{u0}$,
\begin{equation}
\text{\small$
\displaystyle p_{u0}^{ss} = \frac{p_{u}^{eq}\left[1 + e^{-\alpha/\tau}(2\delta-1)
-2\delta e^{-(\alpha-\epsilon)/\tau}\right] + \delta e^{-(\alpha-\epsilon)/\tau}}
{1 + (2\delta-1)(2p_{u}^{eq}-1)\left[e^{-\alpha/\tau}-e^{-(\alpha-\epsilon)/\tau}\right]}$}.
\label{ss}
\end{equation}
The units are chosen such that $k_B T = 1$ and the relaxation time scale of the system, 
$\tau = 1$. With these choices, the steady state probabilities depend on 
the parameters $\alpha$ (cycle time), $\epsilon$ (feedback delay time), $2 U_0$ 
(the level-spacing) and $\delta$ (measurement error). Figure \ref{fig:ss} gives the 
variation of $p_{u0}^{ss}$ as function of $\alpha$ for various combinations 
of $\delta$ and $\epsilon$ for $U_0 = 0.5$. For low values of error, $p_{u0}^{ss}$ lies below the equilibrium value because the feedback is effective in nudging the system to down-state. In the limit $\alpha-\epsilon \rightarrow \infty$, $p_{u0}^{ss} \rightarrow p_{u}^{eq}$, because the system has enough time post feedback to relax back to equilibrium. For a fixed $\delta$, increasing the $\epsilon$ values increase $p_{u0}^{ss}$ since the feedback is less effective with longer delay times. 

Once $p^{ss}_{u0}$ is known, one can use Eq. (\ref{probT}) to find $p_{u}^{ss}(t)$ for $\epsilon < t \le \alpha$. For $0 < t \le \epsilon$, $p_{u}^{ss}(t)$ can be found using the relation $p_{u}^{ss}(t) = \sum_{X',M} P_2(X = u, t | X'; t = 0) \; P_0(X', M)$. It is intriguing to observe that when $\delta$ is non-zero, there exists a specific value of $\epsilon$ for which $p_{u0}^{ss} = p_u^{eq}$ holds true for all $\alpha$. This particular value of $\epsilon$ depends on $\delta$ and $U_0$. The fact that such a combination of parameters exists, which remains independent of $\alpha$, becomes apparent when one equates the expression for $p_{u0}^{ss}$ from Eq. (\ref{ss}) to $p_{u}^{eq}$ (see the Supplement). This result, in turn, implies that, for that particular combination of error and feedback delay time, the system behaves as if it is in equilibrium between the time of measurement and the time of feedback.

Whenever the system's state is switched, the information engine extracts work. Whether it extracts positive or negative work depends on the state of the system at the time of the switch. The demon carries out the switch only when the state is measured to be in the up-state at the beginning of the cycle. Since the measurement is imprecise, this can happen in two different ways: (i) The state of the system is up-state, and the measurement outcome also gives up-state, and (ii) the state of the system is down-state, but the measurement outcome falsely gives up-state. Whether the work extracted is positive or negative in both events 
depends on the system's state during the switch. If the system's state is up-state during the switch, a positive work of $2U_0$ is extracted; if not, a negative work of $-2U_0$ is extracted. Thus, the average work extracted in the steady state by 
the information engine is
\begin{equation}
\begin{aligned}
-\left<W\right> &= p_{u0}^{ss}(1-\delta)\Bigl[2U_0\tilde{p} - 2U_0 (1-\tilde{p})\Bigr] \\
&+ p_{d0}^{ss}\delta \Bigl[2U_0 (1-\tilde{p}') - 2U_0\tilde{p}'\Bigr] \;,
\end{aligned}
\label{work}
\end{equation}
where,
\begin{equation}
\tilde p \equiv e^{-\epsilon/\tau}(1 - p_{u}^{eq}) + p_{u}^{eq}
\end{equation}
is the conditional probability, $P_2(X = u; t = \epsilon^{-}|X' = u; t = 0)$. 
That is, given that the system's state is up at the beginning of the cycle, the probability that it is in up-state just before $t = \epsilon$. Similarly,
\begin{equation} 
\tilde p' \equiv e^{-\epsilon/\tau}p_{u}^{eq} + (1-p_{u}^{eq})
\end{equation}
is the conditional probability, $P_2(X = d, t = \epsilon^{-}|X = d, t = 0)$. 
The two terms on the right-hand side of Eq. (\ref{work}) correspond to the 
events of measurement being correct and measurement being wrong, respectively.

The variations of $-\left<W\right>$ with $\alpha$ for the fixed engine parameters are
given in Fig. \ref{fig:work} for the same set of parameter values as in Fig. \ref{fig:ss}. 
It is seen that the work extracted per cycle increases with the cycle 
time and saturates for large values of $\alpha$ compared to the relaxation time. This is because when measurement errors are minimal, the feedback process
significantly increases the probability of the down-state after feedback. 
If the system has enough time to regain the equilibrium distribution, 
the probability of being found in the up-state in the next measurement increases, leading to larger work extraction. As expected, the work extracted is more for low delay time values and small measurement errors. When delay time becomes of the order of the relaxation time, the work extracted becomes almost zero or negative, indicating that the information gained about the state has been rendered useless. This effect is compounded if the demon is faulty. 

The power of the information engine is the average rate of work extracted and is given by
      \begin{equation}
          \displaystyle \Theta \equiv \frac{-\left<W\right>}{\alpha} 
             = \frac{2U_0 p_{u0}^{ss}(1-\delta)(2\tilde{p}-1) - 2U_0 p_{d0}^{ss} \delta(2\tilde{p}'-1)}{\alpha}
             \label{power}
      \end{equation}
where the numerator is the simplified form of Eq. (\ref{work}).
The other important performance parameter of an engine is its efficiency. 
Although the demon succeeds in rectifying thermal fluctuations to do work, it comes with the cost of information processing. The thermodynamic cost of erasure of memory bits associated with the measurement is given by $k_B T I$, where $I$ is the quantity of information obtained. In the context of an information engine, efficiency is defined as the ratio of average work extracted to the average information processing cost. That is,
\begin{equation}
\eta = \frac{-\left<W\right>}{k_B T\left<I\right>} \;,
\label{eff}
\end{equation}
where $\left<I\right>$ is the average information gathered during a cycle.
If the state of the system at the beginning and end of the cycle were not correlated
(which would be the case if $\alpha - \epsilon \gg 1$), the information cost would 
be determined by the average mutual information \cite{cover1999elements},
\begin{equation}
\displaystyle \left<I_a\right>  = \sum_{M, X} P_0(X,M) \ln{\frac{P(M|X)}{P(M)}} \;,
\end{equation}
where $P(M|X)$ is the probability that the measured state of the system is $M$, 
given that the actual state of the system is $X$ and $P(M)$ is the probability
for the measurement giving $M$. One can use the probabilities given in Table I in the supplement to evaluate $\left<I_a\right>$ and is found to be (see the Supplement),
\begin{equation}
\displaystyle\left<I_a\right> = \Delta \ln{\left(\frac{1-\Delta}{\Delta}\right)} - 
\delta \ln{\left(\frac{1-\delta}{\delta}\right)} + \ln{\left(\frac{1-\delta}{1-\Delta}\right)}
\label{I_a},
\end{equation}
where, $\displaystyle \Delta = p_{u0}^{ss} + \delta - 2p_{u0}^{ss}\;\delta$.

Since the information engine's cycle time is finite, the states at the beginning 
and end of the cycle will be correlated. This implies that the mutual information 
between the state of the system at time $t = n\alpha$ and the measurement outcome 
of the system's state at $t = (n-1)\alpha$ will generally be non-zero. 
This mutual information is given by $I_b (X,M_{pc}) = \ln{\frac{P(X|M_{pc})}{P(X)}}$, 
where $X$ and $M_{pc}$ denote the state of the system in the current cycle 
and the measurement outcome in the previous cycle, respectively. $P(X|M_{pc})$ is the probability 
that the state of the system in the current cycle is $X$, given that the measurement 
outcome of the previous cycle was $M_{pc}$. Averaging $\displaystyle I_b$ over 
the joint distribution, $P(X,M_{pc})$, gives
\begin{equation}
 \left<I_b\right>  = \sum_{M_{pc}, X} P(X,M_{pc})
  \ln{\frac{P(X|M_{pc})}{P(X)}}.
\end{equation}
The details of this calculation, as well as the final form of $\left<I_b\right>$, 
are given in the Supplement. The minimum average cost of running the information 
engine would be proportional to the difference between $\left<I_a\right>$ and 
$\left<I_b\right>$. Assuming $\left<I_a\right>$ to be lager than $\left<I_b\right>$, 
the minimal cost is given by,
\begin{equation}
 k_B T\left<I\right> = k_B T\left(\left<I_a\right> - \left<I_b\right>\right) \;.
    \label{I_c}
\end{equation}
Variation of $\left<I \right>$ with $\alpha$ for various feedback delay times with $U_0 = 0.5$
and $\delta = 0, 0.1$ are shown in Fig. \ref{fig:info}. For a fixed value of $\delta$,
$\left<I \right>$ increases with both $\epsilon$ and $\alpha$. This is primarily 
because $p_{u0}^{ss}$ becomes closer to $0.5$ with increasing delay time as well 
as with increasing cycle time (see Fig. \ref{fig:ss}). This results in more 
information being gathered per measurement. There is a drop in the value of
$\left<I \right>$ with increasing measurement error because the mutual
information between $M$ and $X$ is reduced with a larger error.

Using the definition of efficiency given in Eq. (\ref{eff}) and substituting 
for the average work extracted as well as the information cost found above,
\begin{equation}
\eta = \frac{2U_0 \;p_{u0}^{ss}(1-\delta)(2\tilde{p}-1) - 2U_0 \;p_{d0}^{ss}
    \;\delta(2\tilde{p}'-1)}{k_B T\left<I\right>}\;.
    \label{effic}
\end{equation}
Thus, the power and efficiency of the information engine in the steady state have been found in terms of the four key engine parameters: cycle time ($\alpha$), measurement error ($\delta$), the energy difference between levels ($2 U_0$) and the feedback delay time ($\epsilon$). The parameter space to explore is ample, and in the rest of the paper, some of the most interesting results are presented. Since the dependence of efficiency and work done per cycle on $U_0$ and $\epsilon$ (for infinite cycle time and zero error in measurement) has been discussed in a recent work \cite{kiran3}, the focus here is on the other two design parameters: $\alpha$ and $\delta$.

{\it Working regimes of the engine:}
The work extracted, ($-\left<W\right>$), should be positive for the demon 
to function as an engine. $-\left<W\right> = 0$ gives the boundary surface 
in the four-dimensional parameter space that defines the working regime of 
the engine. Since $\delta$ is a parameter that affects work extraction adversely, 
$-\left<W\right>$ will decrease as $\delta$ increases. The maximum possible
$\delta$ that the demon can tolerate and still extract positive work is given by
(see the Supplement)
\begin{equation}
     \delta_{max} = \frac{p_u^{eq}\left(2\tilde{p}-1\right)}{1-4p_u^{eq}(1-\tilde{p})},
    \label{delmax}
\end{equation}
which is independent of the cycle time, $\alpha$. Figure \ref{fig:delta_max} gives the contour plot of $\delta_{max}$ in the $U_0 - \epsilon$ plane. For $U_0 \rightarrow \infty$, $p_{u}^{eq} \rightarrow 0$ and $\delta_{max} \rightarrow 0$. When $U_0 \rightarrow 0$, $p_{u}^{eq} \rightarrow \frac{1}{2}$ and $\delta_{max} \rightarrow \frac{1}{2}$. 
This suggests that, with a fixed value of $U_0$, precise measurement is crucial for extracting positive work at low temperatures, while it becomes less critical at high temperatures. The values of $U_0$ and $\epsilon$ for which $\delta_{max} < 0$ (dark area at the top right corner in Fig. \ref{fig:delta_max}) correspond to the situation where the demon cannot extract positive work even with no measurement error.

\begin{figure*}[]
    \centering
    \subfloat[]{\includegraphics[width=0.25\textwidth]{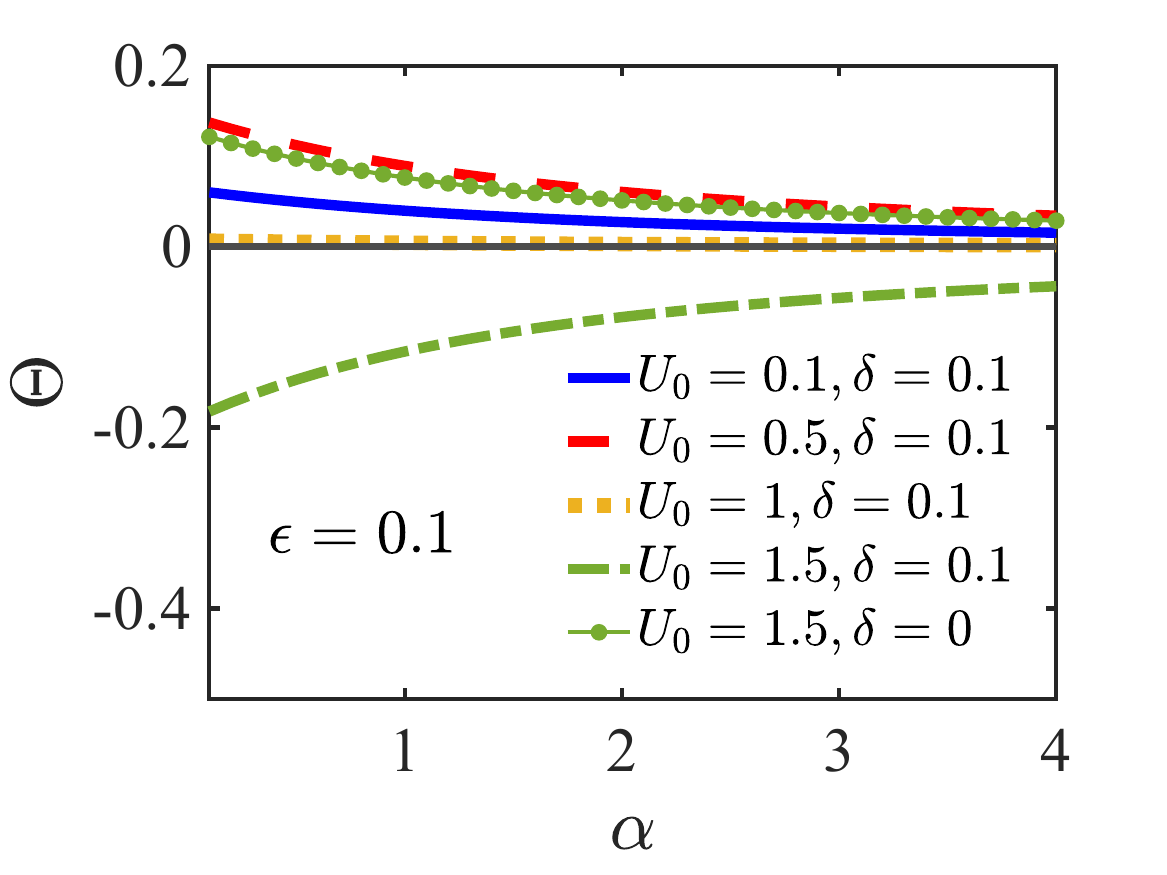}\label{fig:power_sub1}}
    \subfloat[]{\includegraphics[width=0.25\textwidth]{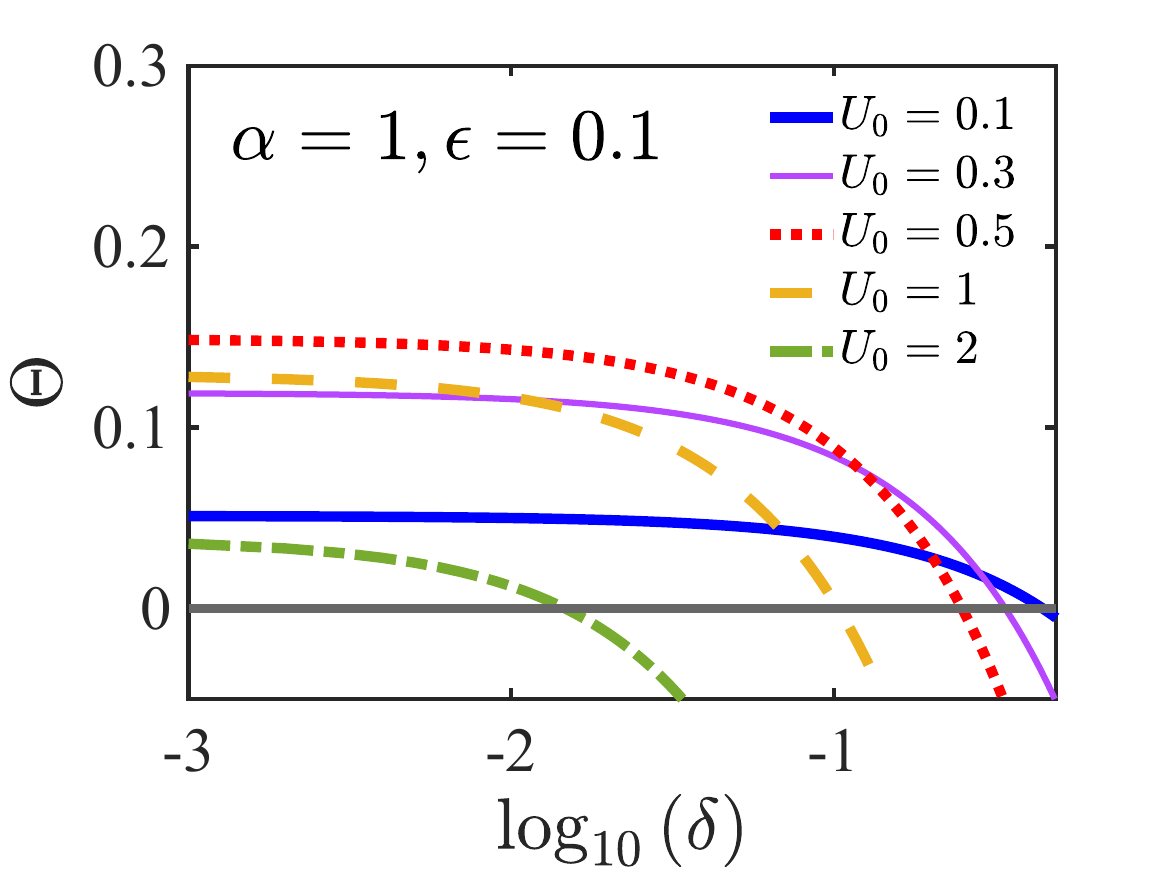}\label{fig:power_sub2}}
    \subfloat[]{\includegraphics[width=0.25\textwidth]{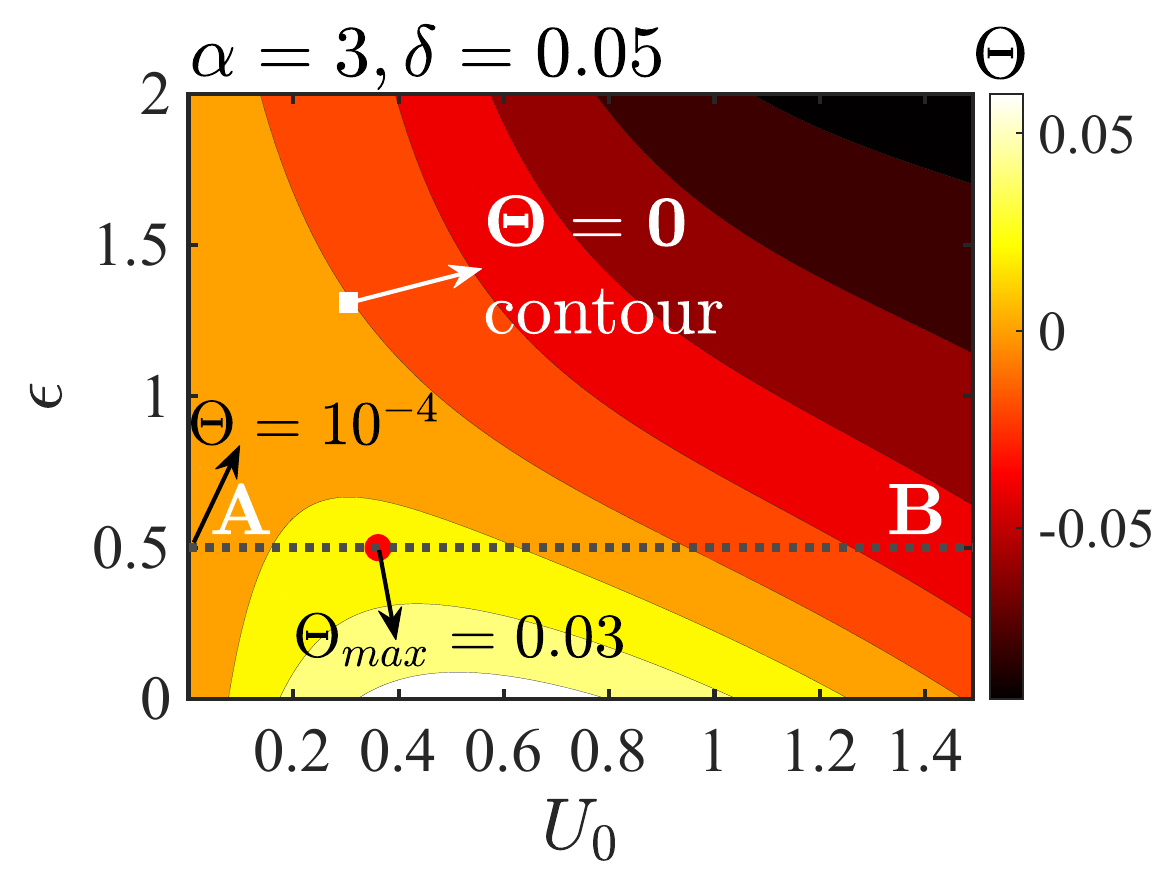}\label{fig:theta2_sub1}}
    \subfloat[]{\includegraphics[width=0.25\textwidth]{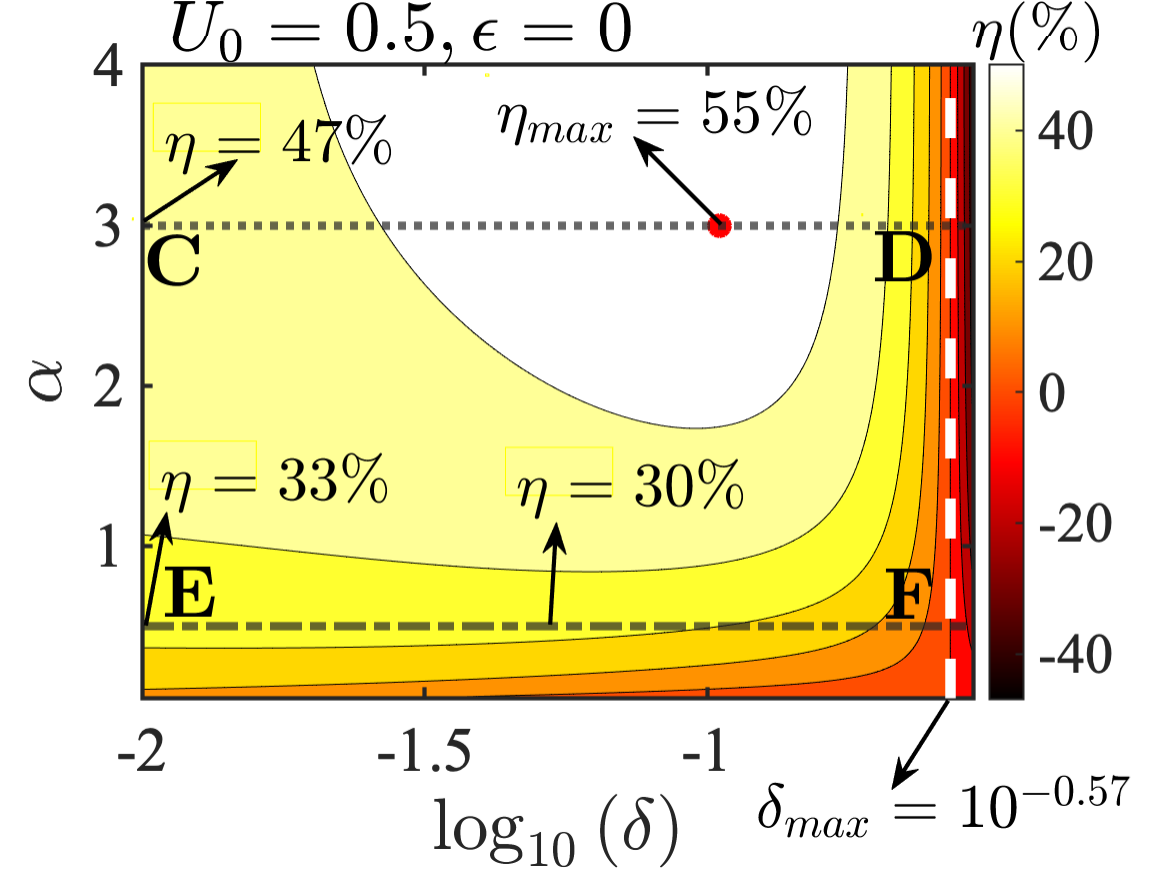}\label{fig:eta_sub0}}
    \captionsetup{justification=centering}
    \caption{Variation of power ($\Theta$) and efficiency ($\eta$) with various engine design parameters are shown. 
    (a) $\Theta$ decreases monotonically with cycle time ($\alpha$) in the regimes where the system
    works as an engine. (b) $\Theta$ reduces with increasing measurement error ($\delta$). At low values
    of level spacing ($2U_0$), the engine can function with larger errors. (c) Variation of $\Theta$ 
    in the $\epsilon - U_0$ plane for the case when $\delta = 0.05$ and $\alpha = 3$. For fixed $\epsilon$, there is a particular value of $U_0$ at which power is a maximum. (d) Dependence of $\eta$ on $\alpha$ and $\delta$
    for $U_0 = 0.5$ and $\epsilon = 0$. For large enough values of $\alpha$, $\eta$ is maximum at an intermediate
    value of $\delta$.}
    \label{fig:main2}
\end{figure*}
{\it Power optimization :} The dependence of power on various engine design 
parameters are shown in Fig. \ref{fig:main2}. Following are some of the key 
observations concerning the power of the engine: \\
(i) For the cases where the device works as an engine ($\Theta > 0$), the
power decreases with increasing cycle time (see Fig. \ref{fig:power_sub1}). 
Even though the work extracted per cycle increases with cycle time 
(see Fig. \ref{fig:work}), the engine's power is compromised for longer 
cycle times because the increase in work grows sub-linearly with cycle time. \\
(ii) For large values of $U_0$ (or equivalently, at low temperatures), even 
a small error in measurement leads to negative values for power. As seen from Fig. \ref{fig:power_sub2}, zero crossing of power happens at smaller $\delta$ values for larger $U_0$. This can be understood as follows: At large $U_0$, the probability of the system being in the up-state will be small at the beginning of the cycle, provided $\alpha-\epsilon \gtrapprox 1$. However, with large errors, there is a likelihood for the measurement to incorrectly record the system as being in the up-state. This, in turn, will lead to negative work extraction during the feedback process because the system will be more likely to transition from down-state to up-state during the state switch. \\
(iii) The power exhibits a non-monotonic dependence on $U_0$. This behavior is evident from the variation of $\Theta$ along the line $AB$ in Fig. \ref{fig:theta2_sub1}, which represents the values of power in the $U_0-\epsilon$ plane for $\alpha = 3$ and $\delta = 0.05$. As $U_0$ approaches zero, the work extracted in each state switch also tends towards zero, resulting in $\Theta \rightarrow 0$. Conversely, at large values of $U_0$, it becomes increasingly improbable for the system to reside in the up-state, leading to a decrease in power. This explains the observed non-monotonic behavior, with the peak power value occurring when $2U_0 \approx 1$. \\
(iv) The range of $\epsilon$ over which the engine extracts positive work 
decreases with increasing $U_0$. This is seen from Fig. \ref{fig:theta2_sub1} where
the $\Theta = 0$ curve in $U_0 - \epsilon$ has negative slopes everywhere. This feature has been observed in an error-free information engine with a cycle time much larger than the relaxation time \cite{kiran3}. However, even with finite cycle time and non-zero error present, the result holds true.

\begin{figure}
    \subfloat[]{\includegraphics[width=0.50\columnwidth]{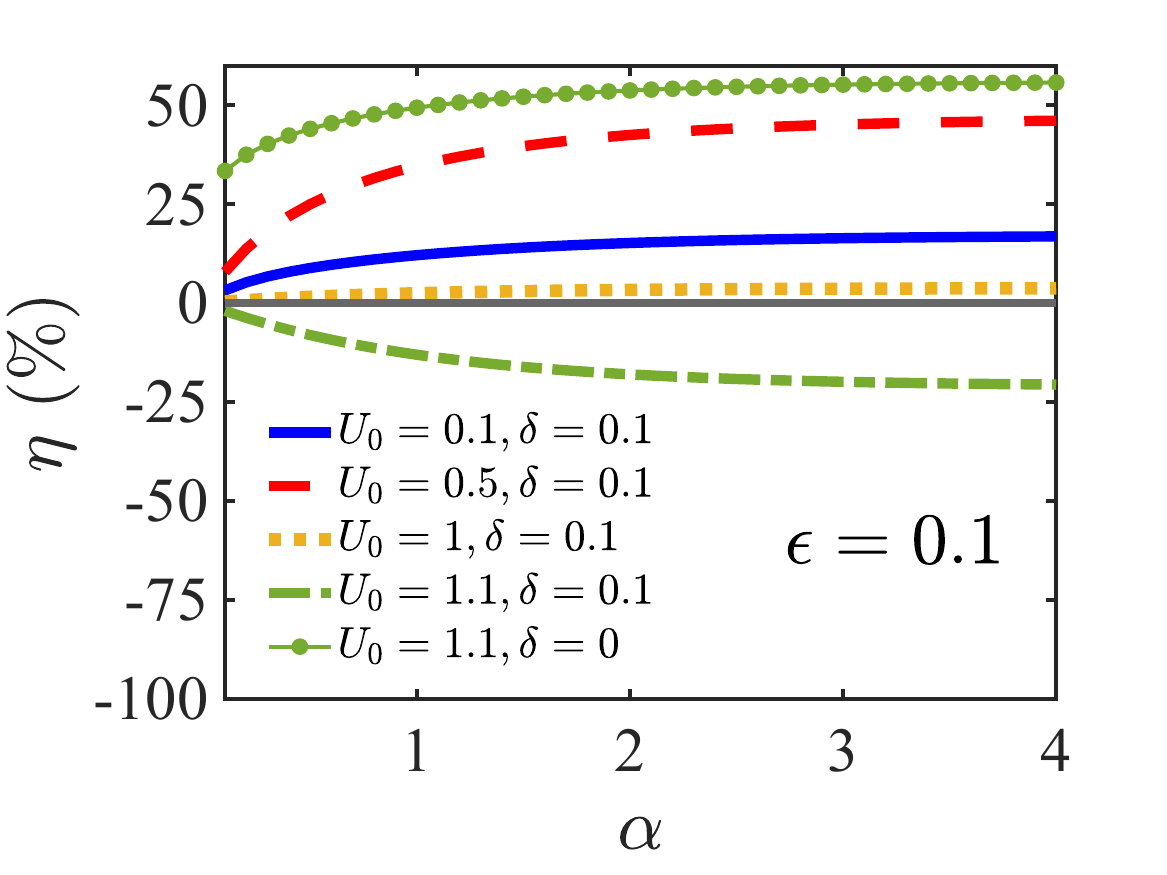}\label{fig:eta_sub1}}%
    \subfloat[]{\includegraphics[width=0.50\columnwidth]{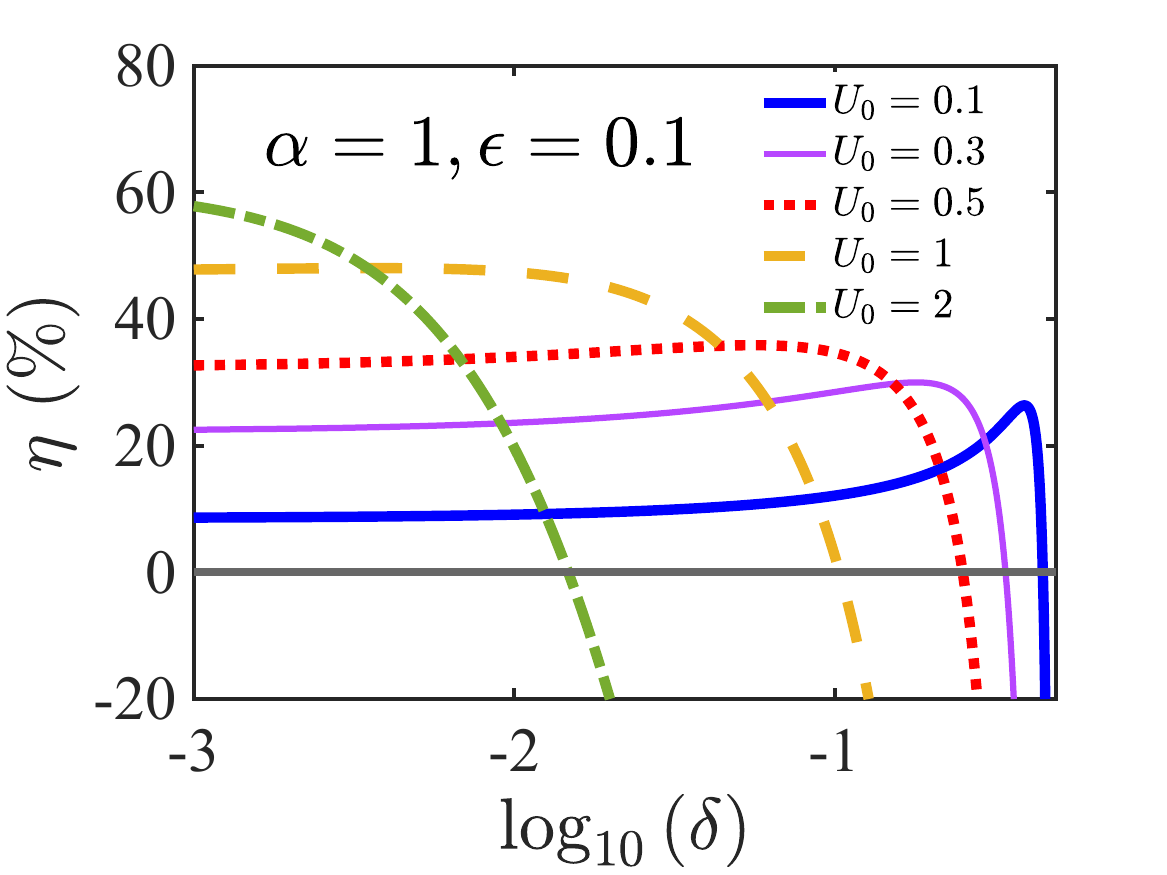}\label{fig:eta_sub2}}\\
    \caption{Dependence of efficiency ($\eta$) on various engine design parameters 
    are shown. (a) $\eta$ shows a monotonic increase with cycle time ($\alpha$) 
    in the regimes where it works as an engine. (b) $\eta$ has a non-monotonic 
    dependence on the measurement error ($\delta$) at low values of level spacing 
    ($2 U_0$), implying that at high temperatures, maximal efficiency is obtained 
    by working with imprecise measurement.}
    \label{fig:eta}
\end{figure}

{\it Efficiency optimization :} The dependence of efficiency on various engine parameters 
is given in Fig. \ref{fig:eta}. Some of the interesting observations concerning the efficiency 
of the engine are: \\
(i) The efficiency increases with cycle time (see Fig. \ref{fig:eta_sub1}). 
Both $-\left<W\right>$ and $\left<I\right>$ increase with $\alpha$ (see discussion related to Fig. \ref{fig:main1}). However, the $-\left<W\right>$ increase is faster, leading to better efficiencies at larger $\alpha$.\\
(ii) For low values of $U_0$, efficiency has a non-monotonic dependence on $\delta$ 
as can be seen for the cases $U_0 = 0.1$, $U_0 = 0.2$, and $U_0 = 0.5$ (blue solid curve, purple thin curve, and red dashed curve respectively in Fig. \ref{fig:eta_sub2}). This means the measurement should have a non-zero error for optimal performance, provided that the power is positive at the chosen error value. This is exemplified by the case of $U_0 = 0.1$ in Fig. \ref{fig:eta_sub2}, where the power is positive for $\delta$ corresponding to the peak efficiency value (approximately $\delta \approx 10^{-0.2}$; refer to Fig. \ref{fig:power_sub2}). \\
(iii) The non-monotonic variation of $\eta$ on $\delta$ depends also on the cycle time as seen in Fig. \ref{fig:eta_sub0}, where
$\eta$ is given in the $\alpha - \delta$ plane for $U_0 = 0.5$ and $\epsilon = 0$.
For example, when $\alpha = 3$ (line $CD$), $\eta$ has a maximum at an intermediate $\delta$ value. But for
$\alpha = 0.5$ (line $EF$), $\eta$ decreases with $\delta$. The advantage of using larger $\alpha$ in this case, with 
better scope for working with less precise measurement values, is limited by the fact that $\Theta$ 
decreases with $\alpha$.\\
(iv) For the same reasons as discussed in the context of power optimization, even a small error can adversely affect efficiency at large values of $U_0$, as seen by the abrupt shift of $\eta$ to negative values when $\delta$ is increased from $0$ to $0.1$ with $\epsilon = 0.1$ and $U_0 = 1.1$ (see Fig. \ref{fig:eta_sub1}).
\begin{figure}
    \subfloat[]{\includegraphics[width=0.50\columnwidth]{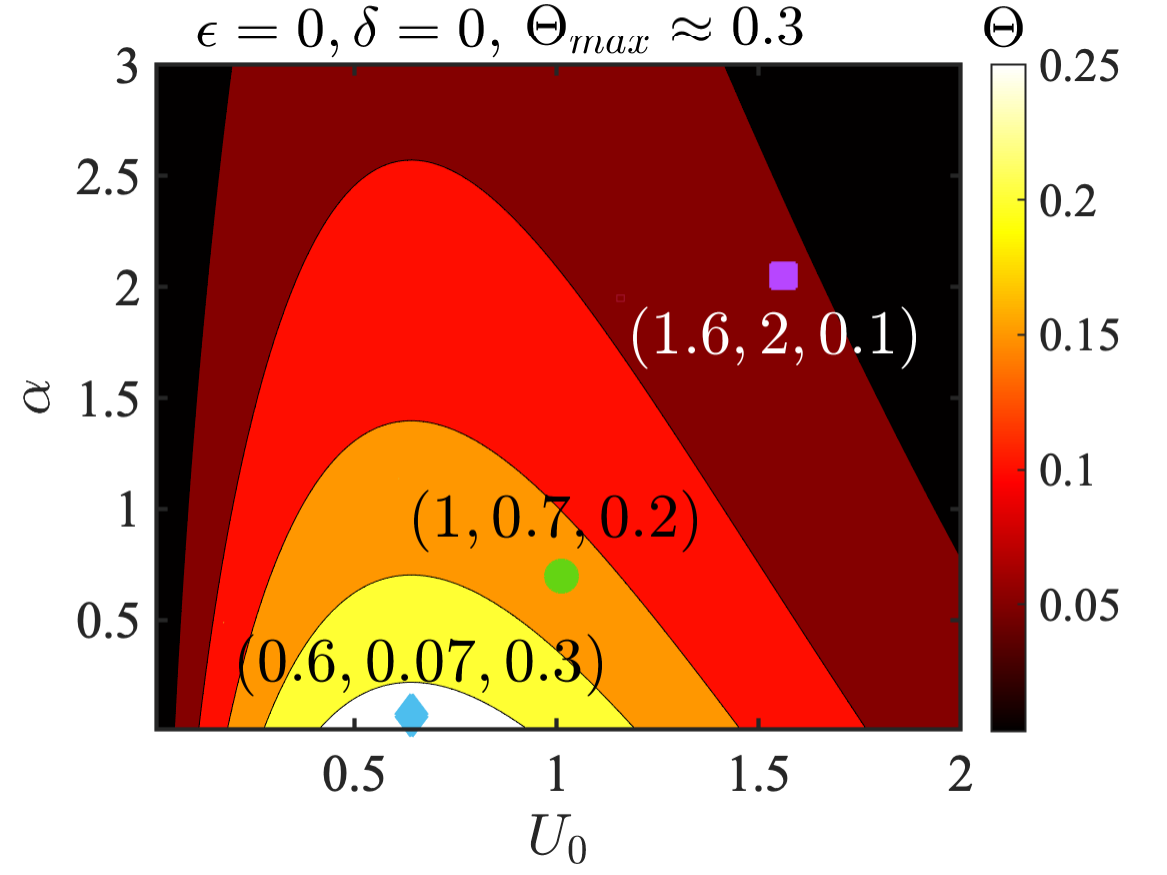}\label{fig:power_sub3}}%
    \subfloat[]{\includegraphics[width=0.50\columnwidth]{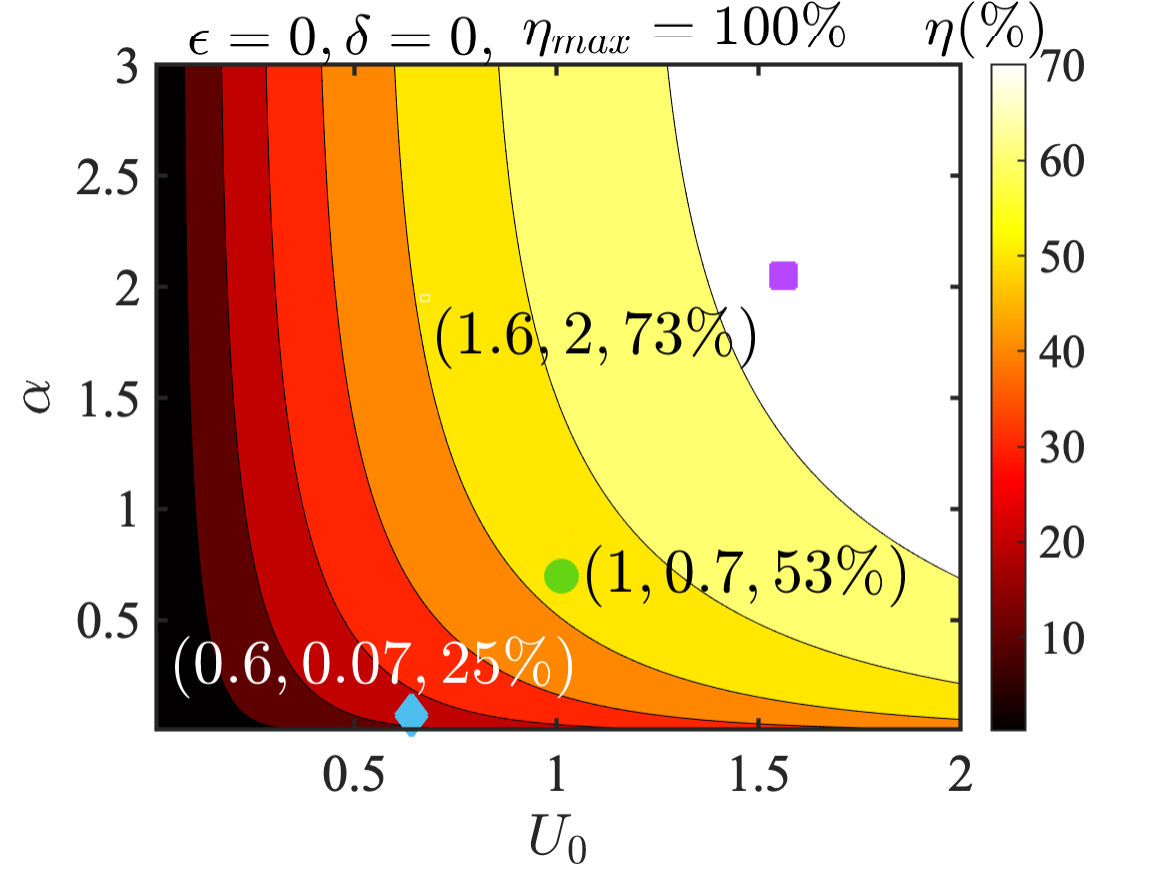}\label{fig:eta_sub3}}\\
    \subfloat[]{\includegraphics[width=0.50\columnwidth]{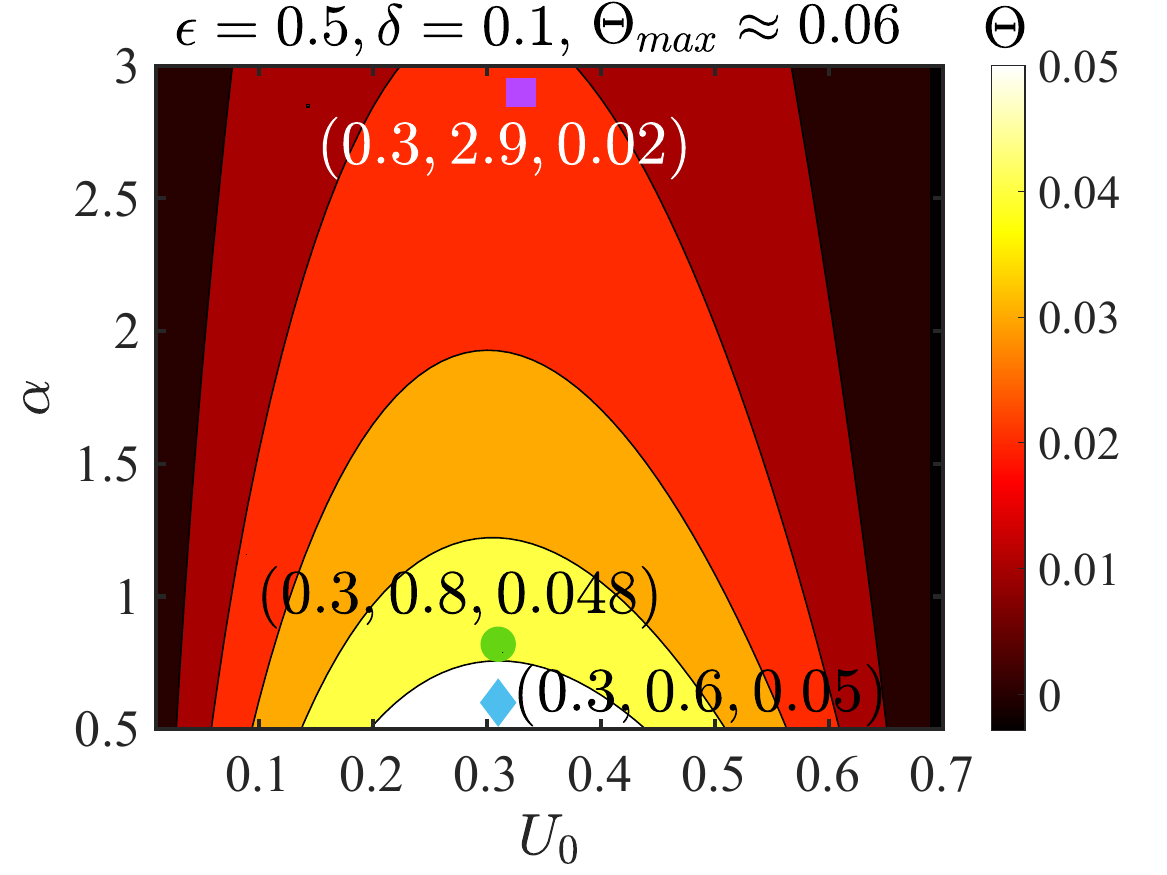}\label{fig:power_sub4}}
    \subfloat[]{\includegraphics[width=0.50\columnwidth]{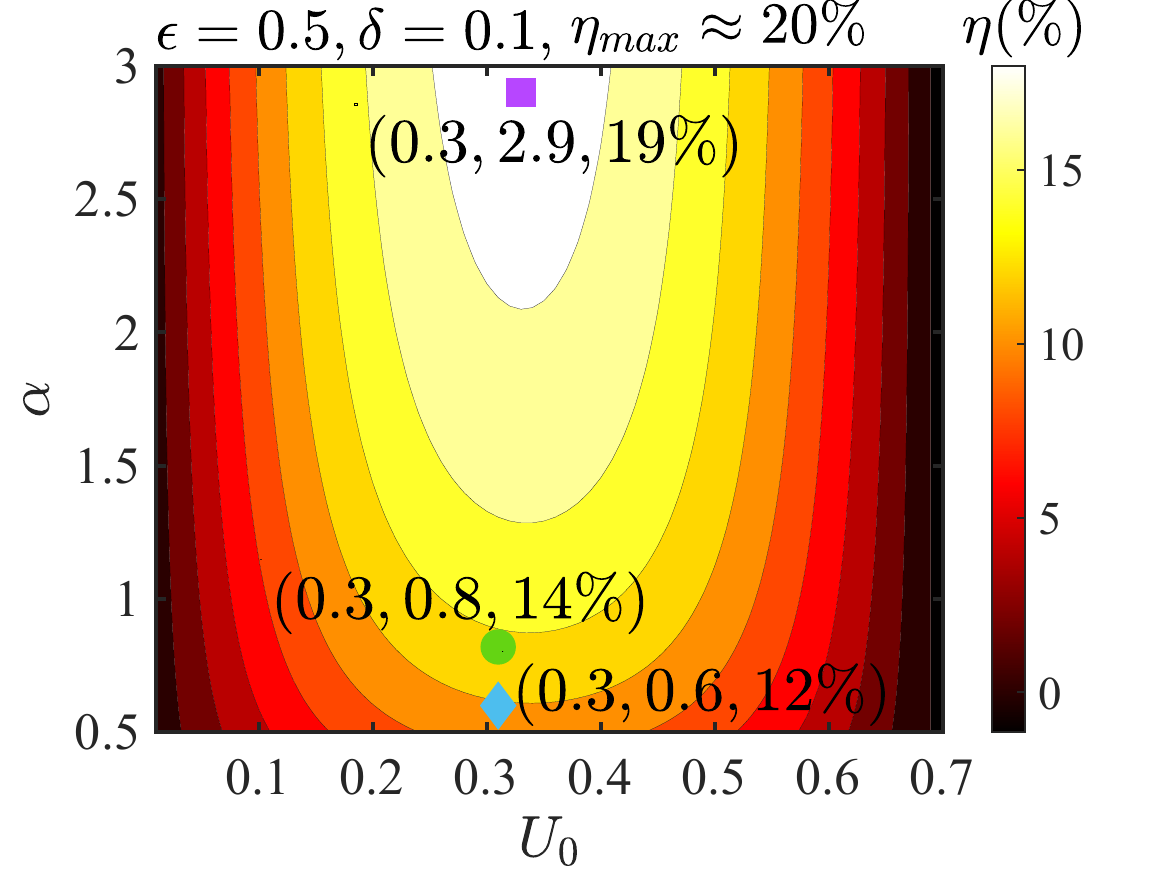}\label{fig:eta_sub4}}
    \caption{Contour plots for power ($\Theta$) and efficiency ($\eta$) in the $U_0 - \alpha$ plane 
    for the case of no feedback delay time ($\epsilon$) and no measurement error ($\delta$) ((a) and (b) respectively) 
    and with $\epsilon = 0.5$ and $\delta = 0.1$ ((c) and (d) respectively). The locations marked
    by violet squares (blue diamonds) give the choice of $\alpha$ and $U_0$ for 
    maximum efficiency (power) with power (efficiency) being at least $20\%$ of its maximum
    value. The green circle indicates the choice of $\alpha$ and $U_0$  for performance 
    that do not compromise either efficiency or power. The values given in brackets are
    the location of the points and the corresponding power (efficiency) value, 
    $(U_0, \alpha, \Theta [\eta])$.}
    \label{fig:optim}
\end{figure}

In an experimental realization of an information engine, the values of $\epsilon$ and $\delta$ tend to be constrained. There is bound to be a limit on the minimum value possible for delay time dictated by the implementation of the feedback process. Similarly, some amount of error is unavoidable in the measurement process. In Fig. \ref{fig:optim}, a comparison is made of the power and efficiency in the $\alpha - U_0$ plane between an ideal scenario where $\epsilon = 0$ and $\delta = 0$ and a more practical situation where $\epsilon = 0.5$ and $\delta = 0.1$ are assumed. As expected, both performance parameters decrease with non-zero $\epsilon$ and $\delta$. The parameters for maximal power with efficiency being at least $20\%$ of the maximum efficiency for both cases are highlighted (blue diamonds in Fig. \ref{fig:optim}). Similarly, the parameters for maximal efficiency with power at least $20\%$ of the maximum power available for both cases are marked (violet squares in Fig. \ref{fig:optim}). Additionally, parameter values for which the power and efficiency are not significantly compromised are indicated (green circles in Fig. \ref{fig:optim}), with both power and efficiency being at least $60\%$ of their maximum values for the corresponding $\epsilon$ and $\delta$ values. With the introduction of error and delay time, one observes appreciable changes in the operating points of the information engine.

{\it Summary:} In this work, a Maxwell's demon based on a two-level system has been studied. To the best of our knowledge, this is the first time that the performance
parameters of an information engine that incorporates feedback delay time, measurement
error, and operating in finite cycle time have been solved analytically. The closed-form 
expression for power and efficiency as a function of level spacing, cycle time, error, 
and delay time allows complete engine characterization. Some of the interesting consequences have been highlighted, including the non-monotonic dependence of efficiency on the error in measurement at high temperatures. A thorough study of the parameter space of the engine will be carried out in future work. The predictions of this work should be verifiable in various experimental systems already implemented
\cite{toyabe2010experimental,koski2014experimental,saha2021maximizing,paneru2018optimal}. Some of the generalizations of the current work that can be studied include a quantum version of the information engine of the type studied here, an information engine design based on a two-level system, but one that involves more than one heat bath, and the design and study of engines based on systems with more than two levels.\\

\noindent
TJ would like to acknowledge the financial support given by SERB-India under 
Grant No: CRG/2020/003646.
\newpage

\onecolumngrid

\appendix
\section{An Imprecise Maxwell's Demon with Feedback Delay: An Exactly Solvable Information Engine Model - Supplemental Material}
Here, we provide the detailed derivations of Eqs. $5, 12, 13$, and $16$ of the main text. Also, the calculation of $\epsilon$ value at which $p_{u0}^{ss} = p_u^{eq}$ for a given $U_0$ and $\delta$ is given.
\section{Calculation of the steady state probability}\label{appendix1}
 Let $p_{u}^{ss}(t) $ be the steady state probability 
of the system to be in the up-state at time $t$. Then, for $\epsilon < t \le \alpha$,
\begin{equation}
     p_{u}^{ss}(t) = \sum_{X'} \biggl[P_2(u; t | X';0) P_0(X',d) \biggr]  
         + \sum_{X', X''} \biggl[ P_2(u; t | X'';\epsilon^{+}) P_1(X''|X') P_0(X',u) \biggr] \; ,
         \label{eq:pssu_append1}
\end{equation}
where $P_0(X, M)$ is the joint probability for $X$ (state) and $M$ (measurement outcome) 
at the time of measurement ($t = 0$). $P_1(X'' | X')$ is the probability for the state 
to be $X''$ at $t = \epsilon^+$ (that is, right after feedback), given that the state was $X'$ at $t = 0$ and the measurement outcome was $M = u$. $P_2(X\; ; t_2\;| X'\; ; t_1)$ is the probability to find the system is in state $X$ at $t = t_2$ given that it was in $X'$ at $t = t_1$ with no interference from the demon in the interval between $t_1$ and $t_2$ and is given by,
\begin{equation}
 P_2(X;t_2|X';t_1) = p^{eq}(X')(1 -  e^{-(t_2 - t_1)/\tau}) 
+ \tilde{\delta}_{X X'} \;e^{-(t_2 - t_1)/\tau}  \;.
\label{P2}
\end{equation}
\noindent
Here $\tilde{\delta}_{X X'}$ is the Kronecker delta function and 
$p^{eq}(X')$ is the equilibrium distribution without feedback. For $X = u$,
\begin{equation}
p^{eq}(X = u) = \frac{e^{-\beta U_0}}{2\cosh{(\beta U_0} )} \equiv p^{eq}_u \;,
\label{eq_prob}
\end{equation}
and $ p^{eq}(X = d) = 1 - p^{eq}_u \equiv  p^{eq}_d$.
Table \ref{ssp} gives the set of various joint and conditional probabilities appearing in 
Eq. (\ref{eq:pssu_append1}), where we have defined $p_{u0}^{ss} \equiv p_{u}^{ss}(t = 0)$ and
$p_{d0}^{ss} \equiv 1 - p_{u0}^{ss}$. Since $p_{u}^{ss}(t)$ is a periodic function 
with period $\alpha$,  $p_{u0}^{ss}$ must be the same as $p_{u}^{ss}(\alpha)$.
Using this fact in Eq. (\ref{eq:pssu_append1}), one gets a self-consistent relation for
$p^{ss}_{u0}$.
\begin{flushleft}
\begin{table*}[b]
\small\addtolength{\tabcolsep}{7pt}
\begin{tabular}{@{}cccccc@{}}
\toprule
\begin{tabular}[c]{@{}c@{}} \hspace{0.6cm} $X'$ \hspace{0.6cm} \end{tabular} &
  \begin{tabular}[c]{@{}c@{}} \hspace{0.6cm} $M$ \hspace{0.6cm} \end{tabular} &
  $P_0(X',M)$ &
  \begin{tabular}[c]{@{}c@{}}\hspace{0.6cm} $X''$ \hspace{0.6cm} \end{tabular} &
  \hspace{0.6cm}$P_1(X''|X')$\hspace{0.6cm} &
  \hspace{0.6cm}$P_2(X = u;t|X'';t=\epsilon^{+})$\hspace{0.6cm} \\
  \hline
   {$u$} &
   {$u$} &
  {$p^{ss}_{u0}.(1-\delta)$} &
  $u$ &
  $p^{eq}_d.(1-e^{-\epsilon/\tau})$ &
  $1-p^{eq}_d.(1-e^{-(t - \epsilon)/\tau})$ \\
 $u$&
  $u$ &
   {$p^{ss}_{u0}.(1-\delta)$} &
  $d$ &
  $1-p^{eq}_d.(1-e^{-\epsilon/\tau})$ &
  $p^{eq}_u.(1-e^{-(t - \epsilon)/\tau})$ \\
  {$d$} &
 {$u$} &
 {$p^{ss}_{d0}.\delta$} &
  $u$ &
  $1-p^{eq}_u.(1-e^{-\epsilon/\tau})$ &
  $1-p^{eq}_d.(1-e^{-(t - \epsilon)/\tau})$ \\
 $d$&
  $u$ &
  $p^{ss}_{d0}.\delta$ &
  $d$ &
  $p^{eq}_u.(1-e^{-\epsilon/\tau})$ &
  $p^{eq}_u.(1-e^{-(t - \epsilon)/\tau})$ \\ 
$d$ &
  $d$ &
  $p^{ss}_{d0}.(1-\delta)$ &
  - &
  - &
  $p^{eq}_u.(1-e^{-t/\tau})$ \\
$u$ &
  $d$ &
  $p^{ss}_{u0}.\delta$ &
  - &
  - &
  $1-p^{eq}_d.(1-e^{-t/\tau})$ \\
  \hline 
\end{tabular}
\caption{The table lists out various joint and conditional probabilities used for the 
calculation of steady state probability $p^{ss}_{u}(t)$ (see Eq. (\ref{eq:pssu_append1})). 
$X'$ is the actual state of the system at $t =0$, $M$ is the state measured by the demon, and $X''$ is the state of the system at $t = \epsilon^+$. 
There is no feedback when $M = d$, and $X''$ becomes a redundant variable. 
The fifth and sixth entries in the last column are $P_2(X = u;t= \alpha |X' = d; t=0)$ and
$P(X = u; t= \alpha |X' = u; t=0)$, respectively.}
\label{ssp}
\end{table*}
\end{flushleft}
\begin{eqnarray}
\displaystyle p^{ss}_{u0} &=& p^{ss}_{u0}(1-\delta)\biggl\{p^{eq}_d.\Bigl(1-e^{-\epsilon/\tau}\Bigr)\left[1-p^{eq}_d.\Bigl(1-e^{-(\alpha-\epsilon)/\tau}\Bigr)\right] \nonumber \\ \nonumber 
&+& \left[1-p^{eq}_d.\Bigl(1-e^{-\epsilon/\tau}\Bigr)\right]p^{eq}_u.\Bigl(1-e^{-(\alpha-\epsilon)/\tau}\Bigr)\biggr\} \\ \nonumber
&+& p^{ss}_{d0}(1-\delta)p^{eq}_u.\Bigl(1-e^{-\alpha/\tau}\Bigr) + p^{ss}_{u0}\delta\Bigl[1-p^{eq}_d.\Bigl(1-e^{-\alpha/\tau}\Bigr)\Bigr]\\ \nonumber
&+& p^{ss}_{d0}\delta\biggl\{\left[1-p^{eq}_u.\Bigl(1-e^{-\epsilon/\tau}\Bigr)\right]\left[1-p^{eq}_d.\Bigl(1-e^{-(\alpha-\epsilon)/\tau}\Bigr)\right]\\  
&+& p^{eq}_u.\Bigl(1-e^{-\epsilon/\tau}\Bigr)p^{eq}_u.\Bigl(1-e^{-(\alpha-\epsilon)/\tau}\Bigr)\biggr\}.
\end{eqnarray}\\
Solving for $p^{ss}_{u0}$ and simplifying, we get,
\begin{equation}
    \displaystyle p^{ss}_{u0} = \frac{p^{eq}_u\left[1 + e^{-\alpha/\tau}\Bigl(2\delta-1\Bigr)-2\delta e^{-(\alpha-\epsilon)/\tau}\right] + \delta e^{-(\alpha-\epsilon)/\tau}}{1 + (2\delta-1)(2p^{eq}_u-1)\left[e^{-\alpha/\tau}-e^{-(\alpha-\epsilon)/\tau}\right]}.
    \label{eq:pssu_append}
\end{equation}
Also, 
\begin{equation}
    \displaystyle p^{ss}_{d0} = 1 - p^{ss}_{u0}.
\end{equation}\\

\section{Value of $\epsilon$ at which $p_{u0}^{ss} = p_u^{eq}$ for a given $U_0$ and $\delta$}
When the measurement error, $\delta$, is not zero, there is a particular value of feedback delay time, $\epsilon$, for which the steady state probability for the system to be in the up-state at the time of measurement will be same as the equilibrium probability ($p_{u0}^{ss} = p_u^{eq}$) independent of the cycle time, $\alpha$. This unique value of $\epsilon$ depends on both $\delta$ and the level spacing, $U_0$. Such a combination of parameters independent of $\alpha$ can be seen from the steady state probability equation by equating it with the equilibrium probability equation.
Considering the condition $p_{u0}^{ss} = p_{u}^{eq}$, one can write the Eq. \ref{eq:pssu_append} as
\begin{equation}
    \displaystyle p^{eq}_u \left[(2\delta-1)(2p^{eq}_u-1)-(2\delta -1)(2p^{eq}_u-1)e^{\epsilon}-(2\delta-1)+2\delta e^{\epsilon}\right] = \delta e^{\epsilon}.
\end{equation}
Solving it for $\epsilon$, one can obtain the following:
\begin{equation}
    \displaystyle \epsilon = \ln{\left\{\frac{2p^{eq}_d p^{eq}_u(2\delta-1)}{(2p^{eq}_u-1)\left[\delta-p^{eq}_u(2\delta-1)\right]}\right\}}.
\end{equation}
\section{Mutual information}\label{appendix2}
\subsection{Calculation of $\left<I_a(X;M)\right>$}
To find the mutual information $\left<I_a(X;M)\right>$ between the variables $X$ and $M$, one has to average all the possible values of $I_a(X,M)$ over all corresponding joint distributions $P(X,M)$. The variables $X$ and $M$ denote the system's actual and measured states at time $t = 0$, respectively. 
The information content is defined as \cite{cover1999elements}
\begin{equation}\label{info_content}
  \displaystyle I_a (X,M) = \ln{\frac{P(M|X)}{P(M)}},\ \   P(M) \neq 0.   
\end{equation}
$P(M|X)$ is defined as the probability that the measured state of the system to be $M$, given that the actual state of the system is $X$. One can obtain the mutual information as
\begin{equation}\label{mutual_info1}
    \displaystyle \left<I_a(X;M)\right> =  \sum_{X,M} P(X,M) \ln{\frac{P(M|X)}{P(M)}}.  
\end{equation}
One can use the table \ref{tab:Ia} to obtain the conditional probabilities $P(M|X)$ corresponding to different combinations of $X$ and $M$ and the marginal probabilities $P(M)$ for the possible measurement outcomes.
\begin{table*}[]
\small\addtolength{\tabcolsep}{10pt}
\begin{tabular}{@{}cccccc@{}}
\toprule
\textbf{$X$} & \textbf{$M$} & \textbf{$P(X,M, t=0)$}    & \textbf{$P(M)$} & \textbf{$P(M|X, t= 0)$} & \textbf{$I_a$} \\ \midrule
u            & u            & $p^{ss}_{u0}(1-\delta)$ & $P(M=u) = p_{u0}^{ss} + \delta - 2p_{u0}^{ss}\delta$ & $1-\delta$ & $\ln{\left(\frac{1-\delta}{p^{ss}_{u0}+\delta-2p^{ss}_{u0}\delta}\right)}$   \\
d            & u            & $(1-p^{ss}_{u0})\delta$     &  & $\delta$ & $\ln{\left(\frac{\delta}{p^{ss}_{u0}+\delta-2p^{ss}_{u0}\delta}\right)}$     \\
u            & d            & $p^{ss}_{u0}\delta$     & $P(M=d) = 1 -  p_{u0}^{ss} - \delta + 2p_{u0}^{ss}\delta$ & $\delta$ & $\ln{\left(\frac{\delta}{1+2p^{ss}_{u0}\delta-p^{ss}_{u0}-\delta}\right)}$   \\
d            & d            & $(1-p^{ss}_{u0})(1-\delta)$ &  & $1-\delta$ & $\ln{\left(\frac{1-\delta}{1+2p^{ss}_{u0}\delta-p^{ss}_{u0}-\delta}\right)}$ \\ \bottomrule
\end{tabular}
\caption{The joint and conditional probabilities required for computing mutual information.}
\label{tab:Ia}
\end{table*}
The marginal probabilities are obtained by the equations
\begin{eqnarray}\label{eq:marg}
      \displaystyle P(M) = \sum_{X} P(X,M), \\  [10pt]
      \displaystyle P(X) = \sum_{M} P(X,M).       
\end{eqnarray}
Using the columns $2$ and $3$ of the table \ref{tab:Ia} and the Eq. \ref{eq:marg}, one can evaluate the marginal probabilities of the possible measurement outcomes as
\begin{eqnarray}
    \displaystyle P(M=u) &=&  P(X=u,M=u) + P(X=d,M=u) \nonumber \\ [5pt]
         \displaystyle   &=&  p^{ss}_{u0}-\delta p^{ss}_{u0} + \delta - \delta p^{ss}_{u0} \nonumber  \\  [10pt]
    \displaystyle P(M=u) &=&  p^{ss}_{u0} + \delta - 2\delta  p^{ss}_{u0} \label{marg:M1} \\     [5pt]
    \displaystyle P(M=d) &=&  1- P(M=u) \label{marg:M2}
\end{eqnarray}
Similarly, using the columns $1$ and $3$ of the  table \ref{tab:Ia}, one can obtain the marginal probabilities of the possible states of the system as
\begin{eqnarray}
    \displaystyle P(X=u) &=&  p^{ss}_{u0} \\  [5pt]
    \displaystyle P(X=d) &=& p^{ss}_{d0}
\end{eqnarray}
The conditional probabilities $P(M|X)$ for the different possible combinations of $X$ and $M$ are given in the column $5$ of the table \ref{tab:Ia}. In addition, the information content, $I_a(X,M)$ calculated using the Eq. \ref{info_content}, is given in the column $6$ of the table \ref{tab:Ia}. Using Eq. \ref{mutual_info1} and the table \ref{tab:Ia}, the mutual information between the variables $X$ and $M$ is obtained as
\begin{eqnarray}
    \displaystyle \left<I_a\right> &=& p^{ss}_{u0}(1-\delta)\ln{\left(\frac{1-\delta}{p^{ss}_{u0} + \delta-2p^{ss}_{u0}\delta}\right)} + 
    p^{ss}_{d0}(1-\delta)\ln{\left(\frac{1-\delta}{1+2p^{ss}_{u0}\delta-p^{ss}_{u0}-\delta}\right)} \nonumber \\ &+& 
    p^{ss}_{u0}\delta \ln{\left(\frac{\delta}{1+2p^{ss}_{u0}\delta-p^{ss}_{u0}-\delta}\right)} + 
    p^{ss}_{d0}\delta \ln{\left(\frac{\delta}{p^{ss}_{u0}+\delta-2p^{ss}_{u0}\delta}\right)}
\end{eqnarray}\\

\subsection{Calculation of $\left<I_b(X_{cc};M_{pc})\right>$}
The mutual information between the variables $X_{cc}$ and $M_{pc}$ is defined as,
\begin{equation}\label{mutual:info2}
     \displaystyle \left<I_b(X_{cc};M_{pc})\right> =  \sum_{X_{cc},M_{pc}} P(X_{cc},M_{pc}) \ln{\frac{P(X_{cc}|M_{pc})}{P(X_{cc})}},  
\end{equation}
where $X_{cc}$ and $M_{pc}$ denote the system's actual state in the current cycle and the measured state of the system in the previous cycle, respectively. $P(X_{cc}, t = \alpha|M_{pc}, t = 0)$ is the probability that the state of the system in the current cycle is $X_{cc}$, given that the outcome of the previous cycle measurement was $M_{pc}$. Note that $cc$ represents \textit{current-cycle} and $pc$ represents \textit{previous-cycle}. The conditional probabilities for the different combinations of $X_{cc}$ and $M_{pc}$ are $P(X_{cc} = u|M_{pc} = d)$, $P(X_{cc} = d|M_{pc} = d)$, $P(X_{cc} = u|M_{pc} = u)$ and $P(X_{cc} = d|M_{pc} = u)$. When $M_{pc} = d$, one can calculate the conditional probabilities using the table \ref{ssp} and the event tree \ref{tree1} given below.
\begin{center}
\resizebox{0.6\textwidth}{!}{
\Tree [.$M_{pc}=d,t=0$ [.$X_{pc}=u,t=0$ [.$X_{cc}=u,t=\alpha$ ] [.$X_{cc}=d,t=\alpha$ ]] [.$X_{pc}=d,t=0$ [.$X_{cc}=d,t=\alpha$ ]
[.$X_{cc}=u,t=\alpha$ ]]]\label{tree1}
}
\end{center}
For example, $P(X_{cc} = u|M_{pc} = d)$ can be calculated as follows: it can be seen from the event tree that when $M_{pc} = d$ at $t=0$, $X_{pc}$ can be either $u$ or $d$. At the beginning of the next cycle at $t = \alpha$, the event of $X_{cc} =u$ can happen through these two independent events. Note that the state of the system is not switched since the measurement result in the previous cycle was $M_{pc}=d$.
Accordingly, one can write
\begin{eqnarray}
   \displaystyle P(X_{cc} = u|M_{pc} = d) &=& \delta P(X_{cc} = u, t = \alpha|X_{pc} = u, t = 0) \nonumber \\ 
    &+& (1-\delta) P(X_{cc} = u, t=\alpha|X_{pc} = d, t = 0).
\end{eqnarray}
Substituting the values of $P(X_{cc} = u|X_{pc} = u)$ and $P(X_{cc} = u|X_{pc} = d)$ from the table \ref{ssp},
    \begin{equation}
          \displaystyle P(X_{cc} = u|M_{pc} = d) = \delta \left[1-p^{eq}_d\left(1-e^{-\alpha/\tau}\right)\right] + (1-\delta)\left[p^{eq}(u)\left(1-e^{-\alpha/\tau}\right)\right].
    \end{equation}
Consequently, one can obtain
\begin{equation}
    \displaystyle P(X_{cc} = d|M_{pc} = d) = 1 - P(X_{cc} = u|M_{pc} = d).
\end{equation}\\
Similarly, the values of the conditional probability can be calculated when $M_{pc} = u$ at $t = 0$, using the table \ref{ssp} and the event tree given below.
\begin{center}
\tikzstyle{level 1}=[level distance=30mm, sibling distance=45mm]
\tikzstyle{level 2}=[level distance=30mm, sibling distance=20mm]
\tikzstyle{level 3}=[level distance=40mm, sibling distance=10mm]
\resizebox{0.6\textwidth}{!}{
\begin{tikzpicture}[grow=right,->,>=angle 60]
      \node{$M_{pc}=u,t=0$}
   child{node{$X_{pc}=u,t=0$}
      child{node{$X_{pc}=u,t=\epsilon$}
      child{node{$X_{cc}=d,t=\alpha$}}
      child{node{$X_{cc}=u,t=\alpha$}}
      }
       child{node{$X_{pc}=d,t=\epsilon$}
      child{node{$X_{cc}=d,t=\alpha$}}
      child{node{$X_{cc}=u,t=\alpha$}}
      }
    }
      child{node{$X_{pc}=d,t=0$}
      child{node{$X_{pc}=u,t=\epsilon$}
      child{node{$X_{cc}=d,t=\alpha$}}
      child{node{$X_{cc}=u,t=\alpha$}}
      }
      child{node{$X_{pc}=d,t=\epsilon$}
      child{node{$X_{cc}=d,t=\alpha$}}
      child{node{$X_{cc}=u,t=\alpha$}}
      }
    };\label{tree2}
 \end{tikzpicture}
}\\
\end{center}
Note that in this case, the evolution of the probabilities depends on the state of the system at $t = \epsilon$ due to the state change. From the event tree it can be observed that, when $M_{pc} = u$ at $t = 0$, $X_{pc}$ can be either $u$ or $d$. With either of these independent initial conditions, $X_{cc}$ at the beginning of the next cycle at $t = \alpha$ depends on the value of $X_{pc}$ at $t=\epsilon$. Again $X_{pc}$ can be either $u$ or $d$ at $t=\epsilon$. For example, consider the case where $M_{pc} = u$ and $X_{pc} = u$ in $t = 0$. Then the probability that the state of the system at $t = \alpha$ is $u$ given that the state was $u$ at $t = 0$ can be written as
\begin{eqnarray}
\displaystyle P(X_{cc} = u, t = \alpha|X_{pc} = u,t=0) &=& P(X_{cc}=u,t = \alpha|X_{pc}=u,t=\epsilon).P(X_{pc}=u,t = \epsilon|X_{pc}=u,t=0) \nonumber \\
&+& P(X_{cc}=u,t = \alpha|X_{pc}=d,t=\epsilon).P(X_{pc}=d,t =  \epsilon|X_{pc}=u,t=0). 
\end{eqnarray}
These probabilities can be obtained from the table \ref{ssp}. Similarly, the expression for $P(X_{cc} = u, t = \alpha|X_{pc} = d,t=0)$ can be obtained from the event tree. Accordingly, one can obtain the conditional probability 
\begin{equation}
    \displaystyle P(X_{cc} = u,|M_{pc} = u) = (1-\delta).P(X_{cc} = u,|X_{pc} = u) + \delta.P(X_{cc} = u|X_{pc} = d).
\end{equation}
Substituting the values of $P(X_{cc} = u,|X_{pc} = u)$ and $P(X_{cc} = u|X_{pc} = d)$ from the table \ref{ssp}
\begin{eqnarray}
    \displaystyle P(X_{cc} = u,|M_{pc} = u) &=& (1-\delta)\biggl\{p^{eq}_d.(1-e^{-\epsilon/\tau}).\left[1-p^{eq}_d.(1-e^{-(\alpha - \epsilon)/\tau})\right] \nonumber\\
     &+& \left[1-p^{eq}_d.(1-e^{-\epsilon/\tau})\right].p^{eq}_u.(1-e^{-(\alpha - \epsilon)/\tau})\biggr\} \nonumber \\
    &+& \delta \biggl\{\left[1-p^{eq}_u.(1-e^{-\epsilon/\tau})\right].\left[1-p^{eq}_d.(1-e^{-(\alpha - \epsilon)/\tau})\right] \nonumber \\ 
    &+& p^{eq}_u.(1-e^{-\epsilon/\tau}).p^{eq}_u.(1-e^{-(\alpha - \epsilon)/\tau})\biggr\}.
    \label{probTX:append}
\end{eqnarray}
Also,
\begin{equation}
    \displaystyle P(X_{cc} = d,|M_{pc} = u) = 1 - \displaystyle P(X_{cc} = u,|M_{pc} = u).
\end{equation}
One can intuitively understand that the marginal probabilities $P(X_{cc} = u)$ and $P(X_{cc} = d)$ are nothing but the steady state probabilities of the corresponding states. In addition, the marginal probabilities $P(M_{pc} = u)$ and $P(M_{pc} = d)$ have already been obtained in Eqs. \ref{marg:M1} and \ref{marg:M2}. One can obtain the joint probabilities using the equation
\begin{equation}
    \displaystyle P(X_{cc},M_{pc}) = P(M_{pc})P(X_{cc}|M_{pc}).
\end{equation}
Substituting the values of $P(X_{cc},M_{pc})$, $P(X_{cc}|M_{pc})$ and $P(X_{cc})$ in Eq. \ref{mutual:info2}, one can obtain the mutual information between $X_{cc}$ and $M_{pc}$
\begin{equation}
     \displaystyle \left<I_b\right> = \Delta c_1\ln{\left(\frac{c_1}{p_{u0}^{ss}}\right)} + \Delta (1-c_1)\ln{\left(\frac{1-c_1}{p_{d0}^{ss}}\right)}
    + (1-\Delta)c_2\ln{\left(\frac{c_2}{p_{u0}^{ss}}\right)} + (1-\Delta)(1-c_2)\ln{\left(\frac{1-c_2}{p_{d0}^{ss}}\right)},
\end{equation}
where
\begin{eqnarray*}
     \Delta &=& p_{u0}^{ss} + \delta-2p_{u0}^{ss}\delta \\
     c_1 &=&  (\gamma - \omega)\left[\mathcal{K}+\omega+\delta(1-\mathcal{K}-\mathcal{L})\right] \\
     c_2 &=& \delta\zeta + (1-\delta)\lambda \\
     \mathcal{K} &=& 1-\tilde{p} \\
     \tilde p &\equiv& e^{-\epsilon/\tau}(1 - p_{u}^{eq}) + p_{u}^{eq}\\
     \mathcal{L} &=& 1-\tilde{p}'\\
     \tilde p' &\equiv& e^{-\epsilon/\tau}p_{u}^{eq} + (1-p_{u}^{eq})\\
     \zeta &=& 1 - p_{d}^{eq}\left[1-e^{-\alpha/\tau}\right] \\
     \lambda &=&  p_{u}^{eq}\left[1-e^{-\alpha/\tau}\right] \\
     \gamma &=& 1 - p_{d}^{eq}\left[1-e^{-(\alpha-\epsilon)/\tau}\right] \\
     \omega &=& p_{u}^{eq}\left[1-e^{-(\alpha-\epsilon)/\tau}\right].
\end{eqnarray*}
The minimal cost of measurement is given by,
\begin{equation}
 k_B T\left<I\right> = k_B T\left(\left<I_a\right> - \left<I_b\right>\right) \;.
    \label{I_c}
\end{equation}
Note that for $\epsilon$ close to $\alpha$ and at large temperatures, $\left<I_b\right>$ can be comparable to or larger than $\left<I_a\right>$. However, this regime is irrelevant to the operation of the demon as an engine because the work extracted will invariably be negative.
\newpage
\section{Upper bounds of $\delta$ and $\epsilon$: Calculation of $\delta_{max}$ and $\epsilon_{max}$}
The average work extracted by the engine is
\begin{equation}
    \displaystyle -\left<W\right> = 2U_0 p_{u}^{ss}(1-\delta)(2\tilde{p}-1) - 2U_0 p_{d}^{ss}\delta(2\tilde{p}'-1),
    \label{eq:work2}
\end{equation}
where,
\begin{equation}
\tilde p \equiv e^{-\epsilon/\tau}(1 - p_{u}^{eq}) + p_{u}^{eq}
\end{equation}
is the conditional probability, $P_2(X = u; t = \epsilon^{-}|X' = u; t = 0)$. 
That is, given that the system's state is up at the beginning of the cycle, 
it is in up-state just before $t = \epsilon$. Similarly,
\begin{equation}
\tilde p' \equiv e^{-\epsilon/\tau}p_{u}^{eq} + (1-p_{u}^{eq})
\end{equation}
is the conditional probability, $P_2(X = d, t = \epsilon^{-}|X = d, t = 0)$.
For the condition $\left<W\right> = 0$,
substituting the values of $p_{u}^{ss}$, $\tilde{p}$ and $\tilde{p}'$, we obtain the equation
\begin{equation}
    \displaystyle b\delta_{max} + c = 0,
    \label{delta_max_solv}
\end{equation}
where
\begin{eqnarray}
    b &=& \left(1-e^{-\alpha/\tau}\right)\left[1-4p_u^{eq}p_d^{eq}(1-e^{-\epsilon/\tau})\right] \\
    c &=& -p_u^{eq}\left(1-e^{-\alpha/\tau}\right)\left[2\left(p_u^{eq}+p_d^{eq}e^{-\epsilon/\tau}\right)-1\right].
\end{eqnarray}\\
Solving the above Eq. \ref{delta_max_solv} for $\delta_{max}$, the upper bound of error tolerance for the engine is obtained as,
\begin{equation}
    \displaystyle \delta_{max} = \frac{p_u^{eq}\left(2\tilde{p}-1\right)}{1-4p_u^{eq}(1-\tilde{p})},
\end{equation}\\
which is independent of the cycle time $\alpha$, but depends on $\epsilon$ and $U_0$.\\
One can solve Eq. \ref{eq:work2} for $\epsilon$, for the same condition $\left<W\right> = 0$, and obtain the upper bound of the feedback delay time for positive work extraction.\\
The upper bound of the feedback delay time for the engine is given by
\begin{equation}
    \displaystyle \epsilon_{max} = -\tau\ln{\left[\frac{p_u^{eq}(2p_u^{eq}-1)-\delta(1-4p_u^{eq}p_d^{eq})}{2(2\delta-1)p_u^{eq}p_d^{eq}}\right]},
\end{equation}
which is also independent of the cycle time $\alpha$, but depends on $\delta$ and $U_0$.
\newline

\twocolumngrid
%

\end{document}